\documentclass[11pt]{article}
\usepackage{amsmath,color}
\usepackage{amsfonts}
\usepackage{amssymb}
\usepackage{mathtools}
\usepackage{enumerate}
\usepackage{graphicx,rotating}
\usepackage{subfig}
\usepackage{float}
\usepackage[font=small, margin=1cm]{caption}
\title{Financial Time Series Analysis and Forecasting\\ with HHT Feature Generation and Machine Learning}
\author{Tim Leung\thanks{Applied Mathematics Department, University of Washington, Seattle WA 98195. Email: {timleung@uw.edu}. Corresponding author.}  \and Theodore Zhao\thanks{Applied Mathematics Department, University of Washington, Seattle WA 98195. Email: {zdzhao16@uw.edu}.}}

\begin{document}

% Title/author/abstract
\maketitle
\bigskip

% Abstract
\abstract{We present the method of complementary ensemble empirical mode decomposition (CEEMD) and Hilbert-Huang transform (HHT) for analyzing nonstationary financial time series. This noise-assisted approach decomposes any time series into a number of intrinsic mode functions, along with the corresponding instantaneous amplitudes and instantaneous frequencies. Different combinations of modes allow us to reconstruct the time series using components of different timescales. We then apply Hilbert spectral analysis to define and compute the associated instantaneous energy-frequency spectrum to illustrate the properties of various timescales embedded in the original time series. Using HHT, we generate a   collection of new features and integrate them into machine learning models, such as regression tree ensemble, support vector machine (SVM), and long short-term memory (LSTM) neural network.  Using empirical financial data, we compare several HHT-enhanced machine learning models in terms of forecasting  performance. }

\maketitle
\newpage
\section{Introduction} Market observations and empirical studies have shown that financial time series such as market indices and asset prices are often driven by multiscale factors, ranging from long-term economic cycles to rapid fluctuations in the short term. This suggests that financial time series are potentially embedded with different timescales. Traditionally, Fourier analysis has been a powerful tool in finding a spectral representation of time series. However, the Fourier bases, by assumption, are stationary and linear. On the other hand, nonstationary and behaviors and nonlinear dynamics are often observed in financial time series. These characteristics can hardly be captured by linear models and call for an adaptive and nonlinear approach for analysis. For decades, methods based on short time Fourier transform and wavelets  have been developed and applied to nonstationary time series, but there are still challenges in capturing nonlinear dynamics, and the often prescribed assumptions make the methods not fully adaptive. This gives rise to the need for an adaptive and nonlinear approach for analysis.

One alternative approach  in adaptive time series analysis is the Hilbert-Huang transform (HHT)  \cite{Huang1998}. The HHT method can decompose any time series into oscillating components with nonstationary amplitudes and frequencies using the empirical mode decomposition (EMD). This fully adaptive method provides a multiscale decomposition for the original time series, which gives richer information about the time series. The instantaneous frequency and instantaneous amplitude of each component are later extracted using the Hilbert transform. The decomposition onto different timescales also and allows for reconstruction up to different resolutions, providing a smoothing and filtering tool that is ideal for noisy financial time series. The method of HHT and its variations have been applied in numerous fields, from engineering to geophysics   \cite{huang2014hilbert}.

There have been several  studies on the variations and alternatives to EMD, including optimization based methods \cite{hou2011adaptive,hou2013data,hou2009variant,huang2013optimization} and noise-assisted approaches \cite{wu2009ensemble,yeh2010complementary}. For financial time series with high level of intrinsic noise, we apply the complementary ensemble empirical mode decomposition (CEEMD) \cite{yeh2010complementary}. Like EMD, CEEMD   decomposes any time series -- stationary or not --  into a number of intrinsic mode functions representing the local characteristics of the time series at different timescales, but the timescale separation is improved by resolving mode mixing in EMD \cite{huang1999new}. The noise-assisted approach is also more robust to intrinsic noise in the data. In previous studies, CEEMD have been found to be useful for forecasting \cite{niu2016novel,tang2015novel} and signal processing \cite{li2015gpr}. More recent application of EMD and the noise-assisted variations includes \cite{saxena2015empirical,nalband2016analysis,nalband2017entropy,nalband2018time}.

Applications of EMD and HHT to finance date back to the work by Huang and co-authors on modeling mortgage rate data \cite{Huang2003}. In previous studies, EMD has been used for financial time series forecasting \cite{emd_forecast,wang2017forecasting} and for examining the correlation between financial time series \cite{nava2018dynamic}. The studies that adopted the full HHT procedure for machine learning appeared in the electrical engineering field \cite{kurbatsky2010using,kurbatskii2011neural,kurbatsky2014}. However, in most of the previous studies using EMD or HHT as forecasting features, the decomposition was pre-implemented on the whole time span, including training and testing data, which is not practical when doing real-time forecasting. Moreover, both EMD and HHT lend information from the past and the future to compute the current values, which can cause information leakage in the forecasting process. On the other hand, extrapolating prediction as in \cite{emd_forecast} is subject to the end effect of EMD, which makes the decomposition error quite high at the end of the time span. It is unknown from previous studies of how to overcome the challenge of end effect  .

In this paper, using   CEEMD and Hilbert spectral analysis, we derive  the instantaneous energy-frequency spectrum associated with   financial time series to  examine the properties of various timescales embedded in the original time series. Important interpretable multiscale outputs from the HHT procedure are recognized as a new set of features, called HHT features. In other words, our method drastically expands the feature set available for machine learning. Compared to previous studies, our main contribution is a novel approach to incorporates HHT features with end effect correction into machine learning methods for extrapolating prediction on financial time series.   We also studied the feature selection problem in using HHT for machine learning.

We experiment the methods using daily data for the S\&P 500 index, CBOE volatility (VIX) index, SPDR gold exchange-traded fund  (GLD), and 10-year treasury yield (TNX). We show that while these time series have contrasting path behaviors (upward trending vs mean-reverting), their instantaneous energy-frequency spectra reveal that they share very similar average frequencies on both long and short timescales. In feature selection part, our comparison analysis identifies the most useful HHT features for prediction. We highlight the improved prediction performance of our method when benchmarked to the prediction using only the original time series. We then discussed the end effect of EMD and proposed a novel framework of extrapolating forecasting with end effect correction. The prediction was implemented on the above-mentioned dataset. Our results are potentially useful not only for prediction but also portfolio construction \cite{Leung_automatica}.

The rest of the paper is structured as follows. First, we  review the methods of EMD and CEEMD and provide the algorithms  for IMF extraction in Sect. \ref{sect-hht}.   The implementation of the methods is illustrated with real financial data   in Sect. \ref{sect-implement}. We also discuss the end effect of the decomposition in Sect. \ref{sect-end} . Next in Sect. \ref{sect-ml}, we introduce HHT features for financial machine  learning. The training and evaluation framework for extrapolating prediction is proposed. We also discuss the incorporation of HHT features into several machine learning models, including regression tree ensemble (RTE), support vector machine (SVM), and especially the structure design of long short-term memory (LSTM) neural network. In Sect. \ref{sect-exp} we discuss the forecasting performance of the HHT-enhanced machine learning (ML) models and feature selection of the HHT features and IMF components. Section \ref{sect-extra} discusses the end effect problem and extrapolating forecasting with end effect correction.

\section{Methodology}\label{sect-hht}
In this section, we present our methodology for processing a financial time series and generating new features for machine learning. The major components are the complementary ensemble empirical mode decomposition (CEEMD) and Hilbert spectral analysis. We also discuss a procedure for time series filtering and reconstruction from the multiscale decomposition. 

\subsection{Empirical Mode Decomposition} \label{sect-emd}
EMD is the first step of our multistage procedure. For any given time series $x(t)$ observed over a period of time $[0,T]$, we decompose it in an iterative way into a finite sequence of oscillating components $c_1(t), \cdots, c_n(t)$, plus a nonoscillatory trend called the residue term $r_n(t)$. Precisely, we have
\begin{equation}\label{eq-emd}
    x(t) = \sum_{j=1}^n c_j(t) + r_n(t).
\end{equation}

To ensure that $c_j(t)$ have the proper oscillatory properties, the concept of intrinsic mode function (IMF) is applied   \cite{Huang1998}. The intrinsic mode functions are real functions in time that admit well-behaved and physically meaningful Hilbert transform. Specifically, each IMF is defined by the following two criteria:
\begin{itemize}
\item No local oscillation: the number of extrema and the number of zero crossings must be equal or at most differ by one.
\item Symmetric: the maxima of the function defined by the upper envelope and the minima defined by the lower envelope must sum up to zero at any time $t\in [0,T]$.
\end{itemize}

The definition of an IMF guarantees pure oscillation while allowing time-dependent frequency and amplitude. Mathematically, we can express am IMF $c(t)$ as 
\begin{equation}\label{eq-af}
    c(t) = a(t)\cos(\theta(t)),
\end{equation} 
where $a(t) \geq 0$ is the instantaneous amplitude, and $\theta(t)$ is the phase function with $\theta'(t) \geq 0$. Such property makes it ready for time-frequency analysis under the Hilbert spectral analysis, which will be discussed in Sect. \ref{sect-hsa}. 

As is standard \cite{Huang1998,rilling2003empirical}, we consider a sifting process that decomposes any time series into a finite set of IMFs that oscillate on different timescales, plus a nonoscillatory residue term. The key idea of the method is as follows: look for the finest oscillation by finding all the local maxima and minima, and then subtract the remaining trend, until the oscillation satisfies the IMF conditions. Each IMF discovered is removed sequentially from the time series until  a nonoscillatory residue remains. The residue is a constant or monotonic function, or has at most one maximum or minimum. The algorithm is summarized as follows:

\begin{itemize}
    \item Initialize the residue term as $r_0(t) = x(t)$ and set $j=1$.
    \item While $r_{j-1}(t)$ does not satisfies the nonoscillatory condition, i.e. $r_{j-1}(t)$ has more than one maximum or minimum, do the following sifting process to extract the oscillation as an IMF denoted by $c_j(t)$.
    \item Initialize the component as $c_j(t) = r_{j-1}(t)$.
    \begin{itemize}
        \item Interpolate the maxima of $c_j(t)$ using a cubic spline as the upper envelope $u(t)$, and interpolate the minima of $c_j(t)$ using a cubic spline as the lower envelope $l(t)$. Compute the mean of the upper and lower envelopes $m(t) = \frac{1}{2}\left( u(t) + l(t) \right)$. 
        \item Iterate $c_j(t) \leftarrow c_j(t) - m(t)$.
        \item Stop when $c_j(t)$ satisfies the criteria of an IMF. Let $c_j(t)$ be the $j$-th component, and iterate $r_j(t) = r_{j-1}(t) - c_j(t)$ and $j\leftarrow j+1$.
    \end{itemize}
    \item Return the IMFs $c_1(t), \cdots, c_n(t)$, and residue $r_n(t)$. 
\end{itemize}
  
The sifting process described above is purely empirical and adaptive with very little assumption on the temporal change, making it ideal for nonstationary time series.

\subsection{Complementary Ensemble Empirical Mode Decomposition}
The EMD algorithm is designed to decompose a time series into multiple components from high frequency to low frequency. However, the phenomenon of mode mixing, first illustrated in \cite{huang1999new}, could occur in some cases. Mode mixing is defined as either one IMF consisting of widely disparate scales, or signals of similar scales reside in several IMF components  \cite{wu2009ensemble}. Such a phenomenon poses potential challenges on the interpretation of the IMFs. This problem is particularly relevant for financial time series when high degree of nonstationarity and noise are observed.

To resolve the mode mixing issue, we consider the ensemble empirical mode decomposition (EEMD) proposed by \cite{wu2009ensemble} for our analysis. EEMD is a noise-assisted signal processing technique that extracts each mode from an ensemble mean computed based on $N$ trials. In each trial $i$, an i.i.d. white noise $w_i$ with a zero mean and finite variance $\sigma$ is added to the original time series $x(t)$, and $x(t) + w_i(t)$ is referred as the  \emph{signal}  in this trial. The original EMD algorithm is then applied to the signal, outputting the IMFs $c_{ij}(t),\ j = 1,\cdots,n$, and the residual term $r_{in}(t)$. Finally, the ensemble mean of the IMFs and residual terms across all the $N$ trails is regarded as the true mode extraction. The resulting ensemble mean of the IMF components are given by
\begin{equation}
    c_j(t) = \frac{1}{N}\sum_{i=1}^N c_{ij}(t),  \quad r_n = \frac{1}{N}\sum_{i=1}^N r_{in}(t).
\end{equation}
Using \eqref{eq-emd}, we can write 
\begin{equation}
    \sum_{j=1}^n c_{ij}(t) + r_{in}(t) = x(t) + w_i(t).
\end{equation}
Then, the ensemble mean of the IMF components sums up to
\begin{align}
    \sum_{j=1}^n c_j(t) + r_n(t)    &= \frac{1}{N}\sum_{i=1}^N (x(t) + w_i(t))= x(t) + \frac{1}{N}\sum_{i=1}^N w_i(t)\,.\notag
\end{align}

Therefore, the components from the ensemble mean provide a decomposition which converges to the original time series $x(t)$ almost surely at the rate of $\mathcal{O}(\frac{1}{\sqrt{N}})$. As such, a large ensemble size is typically desired if possible. Implementing EMD with a large number of trials can be potentially prohibitive in terms of computational cost and speed. To address this issue, an alternative called the complementary ensemble empirical mode decomposition (CEEMD) method is introduced \cite{yeh2010complementary}. CEEMD expands the ensemble by adding a complementary negative noise $-w_i(t)$ for each trial, expanding the total ensemble size to $2N$. Again, the components from the ensemble mean   sum up to equal the original time series:

\begin{equation}\label{eq-ceemd}
     \sum_{j=1}^n c_j(t) + r_n(t) = x(t) + \frac{1}{2N}\sum_{i=1}^N (w_i(t) - w_i(t))=x(t).
\end{equation}
This holds regardless of the choice of $N$, thus reducing the need to have a very large ensemble size. Besides reducing mode mixing, both EEMD and CEEMD are also more robust to intrinsic noise in the original data, as these methods automatically average out extra independent noises in the process. 

 Due to the high non-stationarity and noise level in financial time series, we consider using CEEMD as the noise-assisted decomposition to resolve the mode mixing problem. For the rest of the paper, we use CEEMD to extract the IMF components from time series. Further HHT features such as instantaneous frequency and amplitude are also derived from the CEEMD result.

\subsection{Hilbert Spectrum}\label{sect-hsa}
By definition, an IMF lends itself to the second stage of HHT, Hilbert spectral analysis. An oscillating real-valued function can be viewed as the projection of an orbit on the complex plane onto the real axis. For any function in time $X(t)$, the Hilbert transform is given by
\begin{equation}\label{hilbert}
    Y(t) = \mathcal{H}[X](t) := \frac{1}{\pi} \int_{-\infty}^{+\infty} \frac{X(s)}{t-s}\, ds,
\end{equation}
where the improper integral is defined as the Cauchy principle value. The transform exists for any function in $L^p$ \cite{titchmarsh1948introduction}. As a result, $Y(t)$ provides the complementary imaginary part of $X(t)$ to form an analytic function in the upper half-plane defined by
\begin{equation}
    Z(t) = X(t) + i Y(t) = a(t) e^{i\theta(t)},
\end{equation}
where
\begin{equation}\label{ampEq}
    a(t) = \|Z(t)\| = \sqrt{X^2(t) + Y^2(t)}
\end{equation}
and 
\begin{equation}\label{phaseEq}
    \theta(t) = \arg Z(t) = \arctan(\frac{Y(t)}{X(t)}).
\end{equation}

As is standard \cite{bedrosian,nuttall}, for a function of the form in\eqref{eq-af}, if the amplitude $a(t)$ and the frequency $\theta'(t)$ are slow modulations, the Hilbert transform will give a $\pi/2$ shift to the phase $\theta(t)$. Therefore, the $a(t)$ given by \eqref{ampEq} is exactly the instantaneous amplitude, and the $\theta(t)$ given by \eqref{phaseEq} is exactly the instantaneous phase function. The instantaneous frequency is then defined as the $2\pi$-standardized rate of change of the phase function, that is,
\begin{equation}\label{freqEq}
    f(t) = \frac{1}{2\pi}\dot{\theta}(t) = \frac{1}{2\pi}\frac{d}{dt}\left( \arctan(\frac{Y(t)}{X(t)}) \right).
\end{equation}

By the definition of IMF, it has a well-defined Hilbert transform, which means that $a(t)$ and $f(t)$ computed reflect the  amplitude and frequency of the oscillation, thus the Hilbert transform is physically meaningful.  Applying Hilbert transform to each of the IMF components individually yields  a sequence of analytic signals \cite{Huang1998}:
\begin{equation}\label{eq_complex_imf}
    c_j(t) + i \mathcal{H}[c_j](t) = c_j(t) + i \hat{c}_j(t) = a_j(t) e^{i\theta_j(t)},
\end{equation}
for $j = 1, \cdots, n$. We refer to the $c_j(t) + i\hat{c}_j(t),\ j=1,\cdots,n$ as the complex IMFs, where the IMF components are the real parts, and the corresponding Hilbert transforms serve as the imaginary parts. In turn, the original time series can be represented as
\begin{equation}\label{eq-hs}
    x(t) = \mathfrak{Re}\ \sum_{j=1}^n a_j(t) e^{i \int^t 2\pi f_j(s) ds} + r_n(t).
\end{equation}
This decomposition can be seen as a sparse spectral representation of the time series with time-varying amplitude and frequency. In other words, each IMF represents a generalized Fourier expansion that are suitable for nonlinear and nonstationary financial time series.  
In summary, the procedure generates a series of complex functions that are analytic in time, along with their associated instantaneous amplitudes and instantaneous frequencies. These components capture different time scales and resolutions embedded in the time series and are used for time series filtering and reconstruction.

Lastly, the Hilbert spectrum is defined by
\begin{equation}
   H(f,t) = \sum_{j=1}^n H_j(f,t),  \quad \mbox{where}\ {\displaystyle H_{j}(f ,t)={\begin{cases}a_{j}(t),&f =f_{j}(t),\\0,&{\text{otherwise.}}\end{cases}}}
\end{equation}
The instantaneous energy of the $j$-th component is defined as
\begin{equation}
E_j(t) = |a_j(t)|^2.
\end{equation}
We examine the behavior of the Hilbert spectrum through the pair $(f_j(t),E_{j}(t))$ (see Fig. \ref{fig:mode_spec}). The Hilbert spectrum will also be incorporated into machine learning models for time series forecasting.

%\section{Key Features of Model}\label{sect-keyfeat}
%There are many properties of financial time series that makes it different from other applications, and lead to the key features of our model. Firstly, 

\subsection{Time Series Filtering and Reconstruction}\label{sect-recon}
Financial time series often exhibits characteristics on different timescales, from long-term economic cycles to daily rapid fluctuations. Moreover, financial data can be noisy, so we used a noise-added approach to make the decomposition more robust. CEEMD decomposes the time series onto different timescales, which sum up to the original time series.
 
By construction, the iterative sifting process identifies the finest structures, and then extract longer and longer scales. As a result, the first few components have higher frequencies which are more noisy, and the last few components have lower frequencies representing long term structures. Considering the noisy and complex nature of financial time series, such scale separation is valuable. In fact, EMD can be used as a filter for time series  \cite{Huang2003}. In the reconstruction of the original time series using the IMF components, we can choose a subset of modes as a filter for desired information. By removing the first few high frequency components, we create a low-pass filter; that is,
\begin{equation}
    x_L^{(m)}(t) = \sum_{j=m}^n c_j(t) + r_n(t). \label{lowpass}
\end{equation}
This reconstruction using only the last few components can serve as a smoothing of the time series. Similarly, we can also build a high-pass filter with
\begin{equation}
    x_H^{(m)}(t) = \sum_{j=1}^m c_j(t),
\end{equation}
which captures the high-frequency local behaviors, and can also be used to estimate the noise level or volatility.

\subsection{Implementation with Financial Data}\label{sect-implement}
In this section, we use four financial time series of different type and in different sectors for implementation, namely the S\&P 500 index, the CBOE volatility index VIX, the gold price exchange-traded fund (ETF) GLD, and the 10-year treasury yield (TNX), over a 10-year period from 4/1/2010 to 3/31/2020. For S\&P 500 and GLD, let $s(t)$ be the value of the time series and $x(t) := \log(s(t))$ be its log price. For VIX and 10-year treasury rate, let $x(t)$ be the time series itself. In each case, CEEMD is applied to $x(t)$ to extract the IMFs and residual from the time series, the Hilbert spectrum of which is later computed with HHT. To reduce the error in instantaneous frequency computation, we have used the robust locally weighted scatterplot smoothing (robust LOWESS)  \cite{cleveland1979robust}. The energy-frequency spectrum is then shown to display the instantaneous values of $(f_j(t),E_{j}(t))$.  

\begin{figure}[t]
  \centering  %\includegraphics[trim = 1.1cm   2.4cm  1.5cm  1.5cm, clip, width=3.48in]{eemdSP500.eps}\\
    \includegraphics[trim = 1cm   1.8cm  1.5cm  1.5cm, clip, width=3.2in]{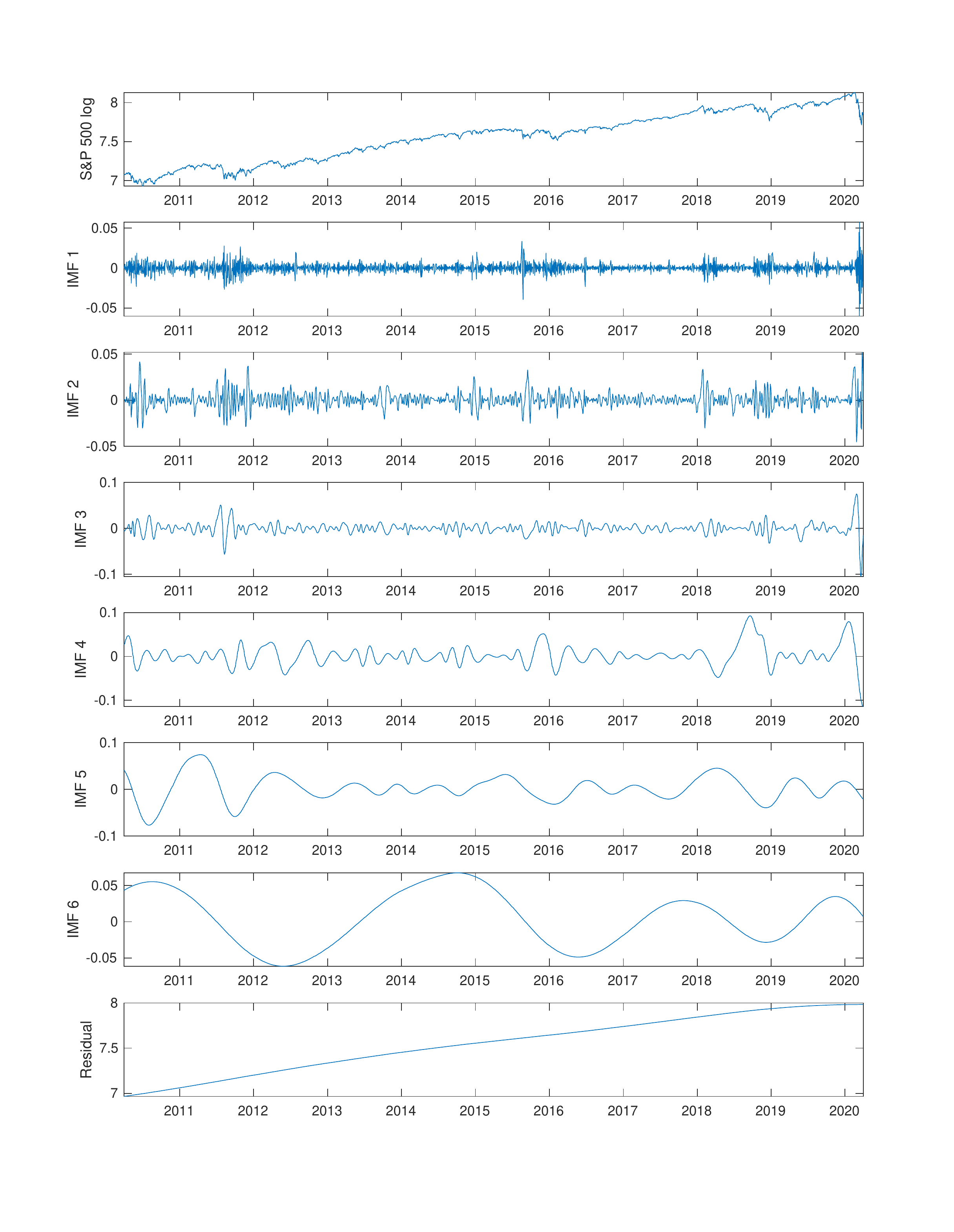}
    \includegraphics[trim = 1cm   1.8cm  1.5cm  1.5cm, clip, width=3.2in]{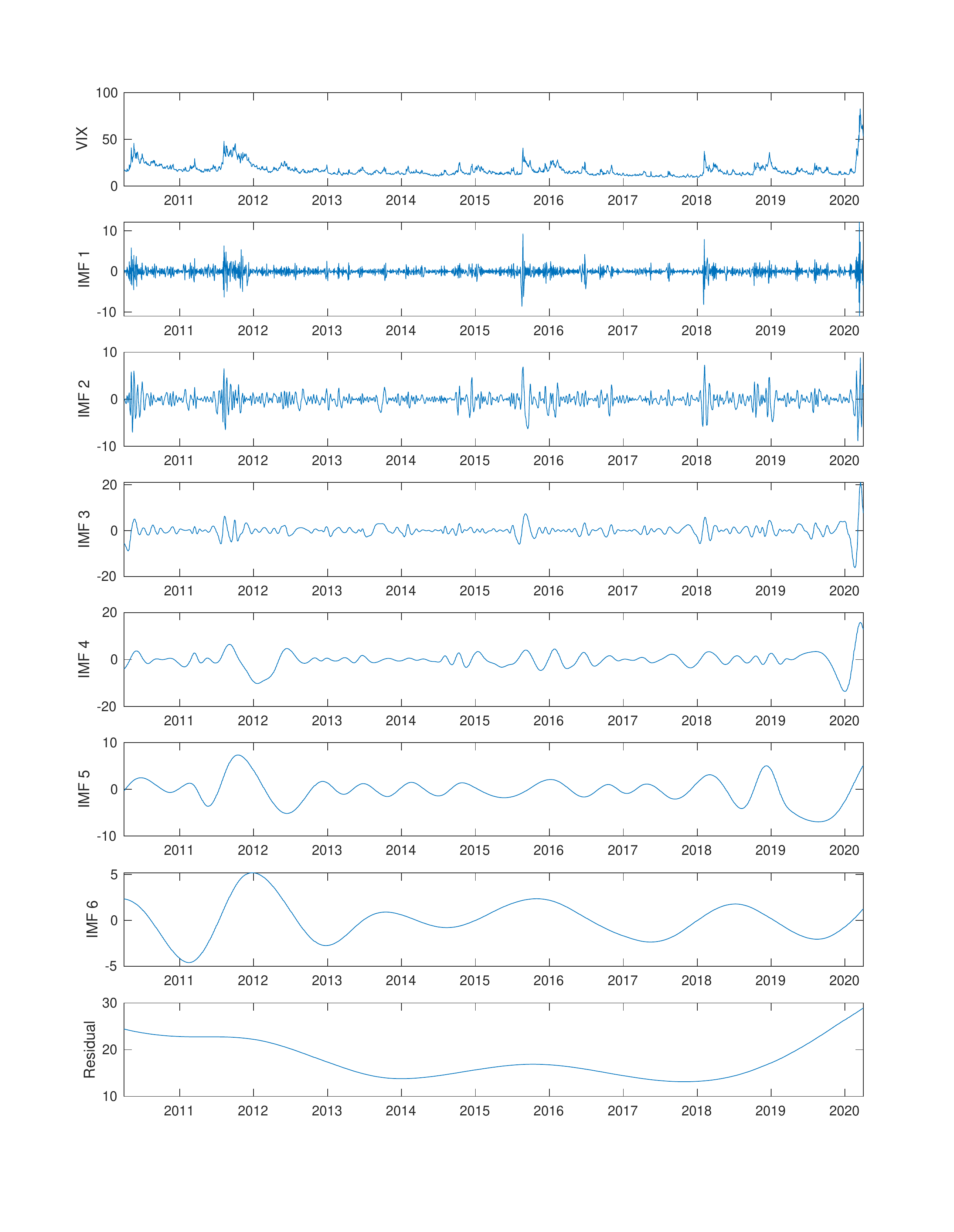}\\
     \includegraphics[trim = 1cm   1.8cm  1.5cm  1.5cm, clip, width=3.2in]{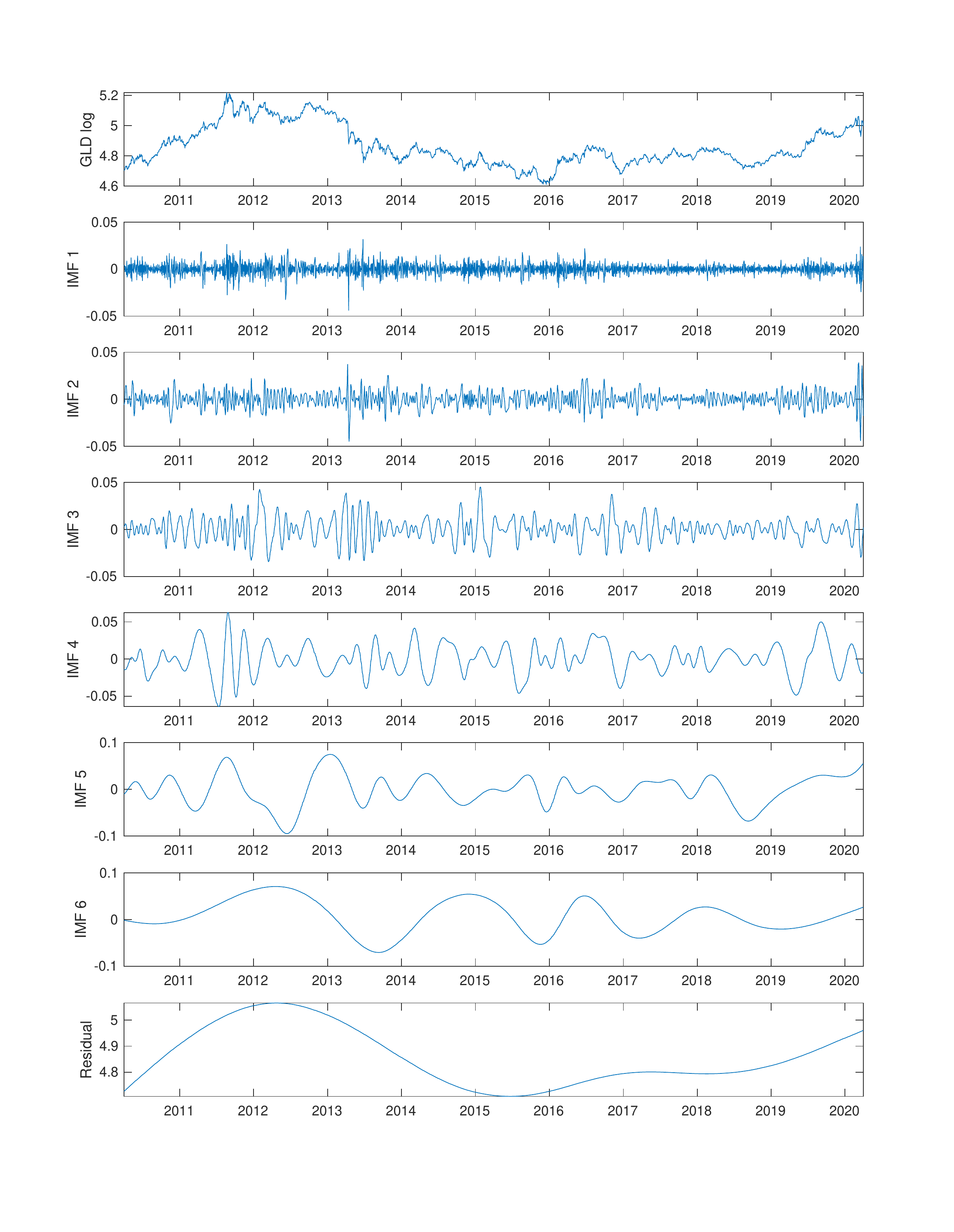}
     \includegraphics[trim = 1cm   1.8cm  1.5cm  1.5cm, clip, width=3.2in]{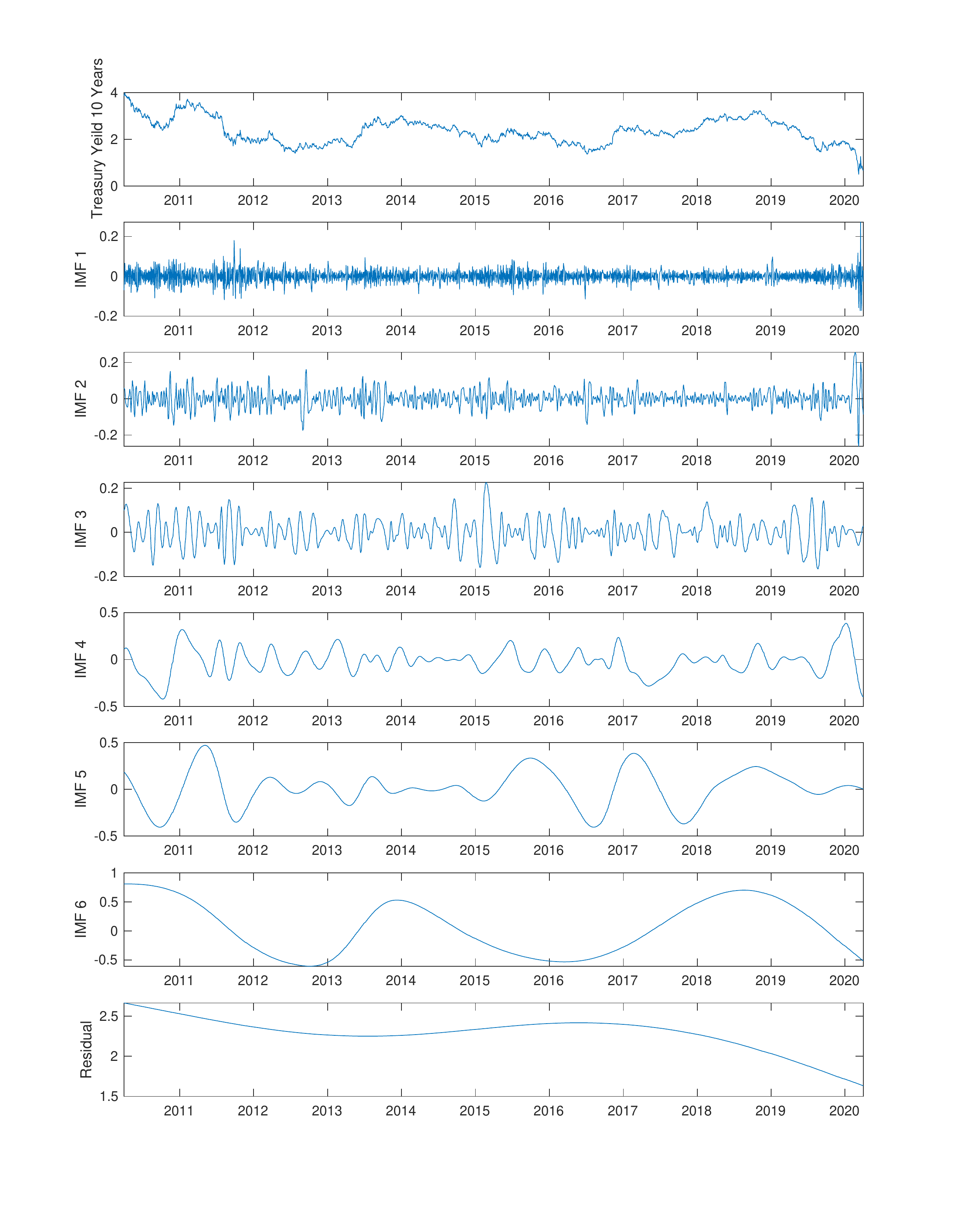}
    \caption{\small{IMFs and residual terms extracted from CEEMD of S\&P 500   (top left), VIX (top right), GLD   (bottom left), and 10-year treasury yield (bottom right), from 4/1/2010 to 3/31/2020. The top row in each plot shows the original time series. The second to last but one rows show the IMF modes of the corresponding time series. The bottom row of each plot shows the residual term of the time series.}}\label{fig_eemd}
\end{figure}

In Fig. \ref{fig_eemd}, we implement CEEMD using the 10-year historical data for the S\&P500, GLD, VIX, and 10-year treasury yield. The top row in each plot shows the original time series $x(t)$, followed by the IMFs with decreasing frequencies, ended with the residual term showing the overall trend. The modes on different timescales can indicate different behavior of the time series. Taking the recent COVID-19 event for example, when S\&P 500 began the sharp decline from 2/21/2020 on,   modes 3, 4, 5 and 6 also show  notable peak around the time period, but on earlier and earlier dates (2/25/2020, 1/21/2020, 12/26/2019, and 11/15/2019). This indicates that a mode on a longer timescale signals trends earlier than other modes. The opposite behavior is observed for VIX -- the lower-frequency modes (modes 3 to 6) exhibit a steep increase towards the day when VIX spikes.

\begin{figure}[ht]
   \centering
    \includegraphics[trim = 0.65cm   0.5cm   1.38cm  0.5cm, clip, width=5.85in]{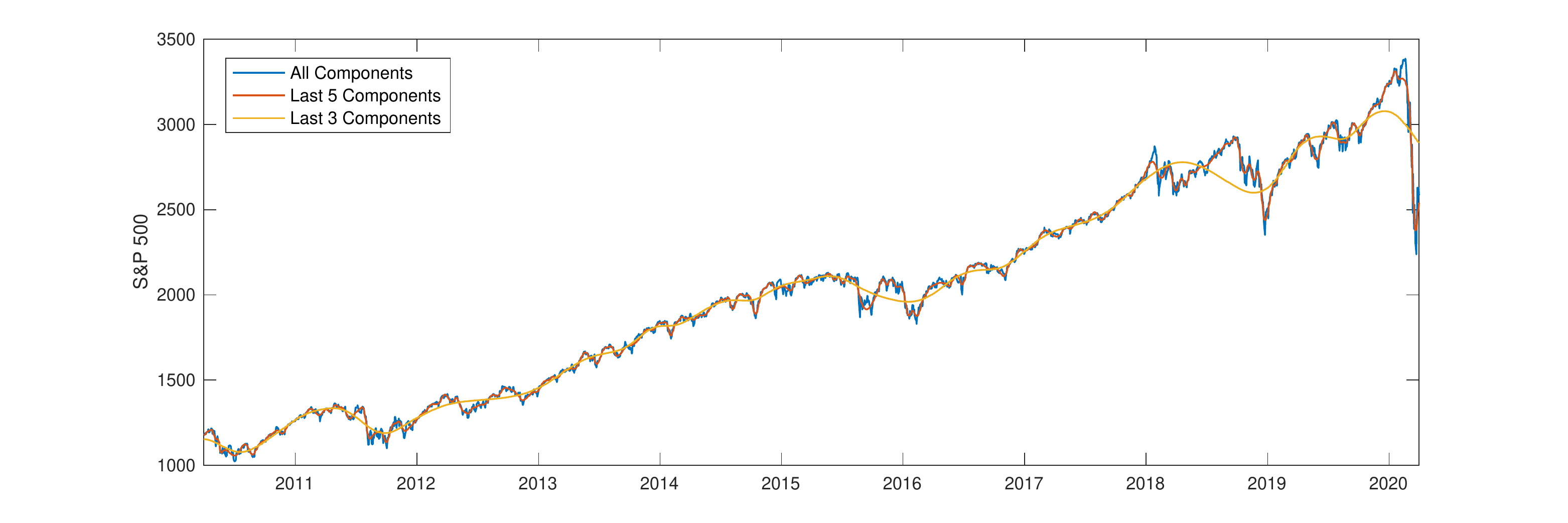} \\
        \includegraphics[trim = 0.65cm   0.5cm    1.38cm  0.5cm, clip, width=5.85in]{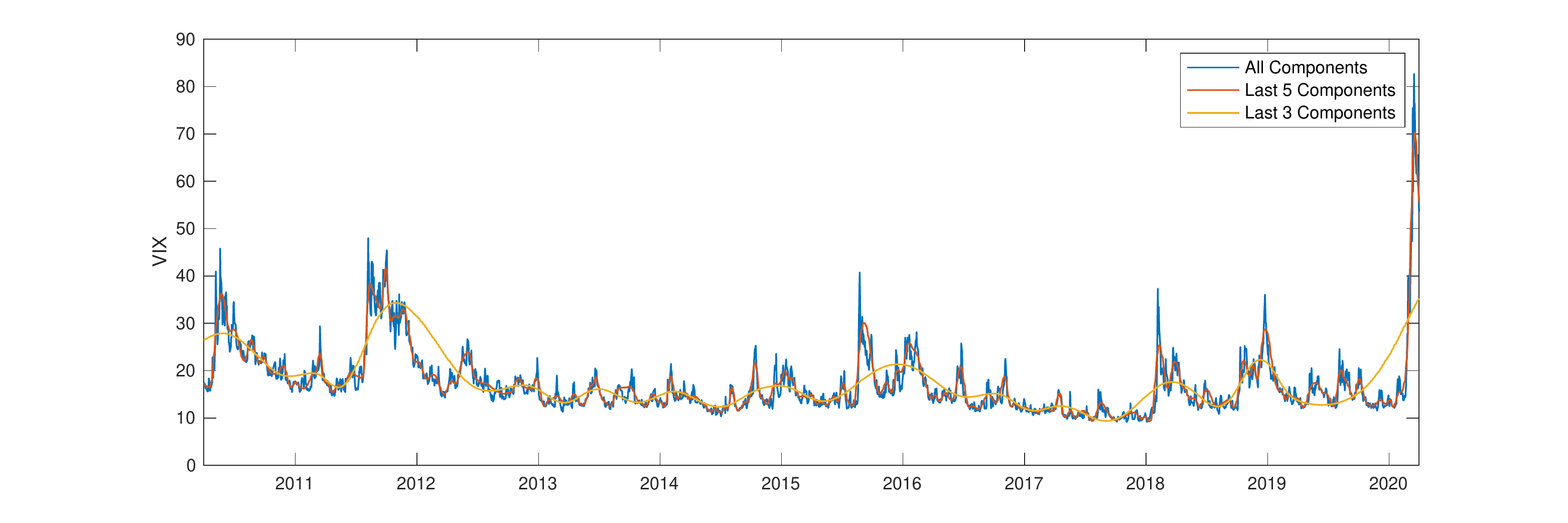} \\
        \includegraphics[trim = 0.65cm   0.5cm    1.38cm  0.5cm, clip, width=5.85in]{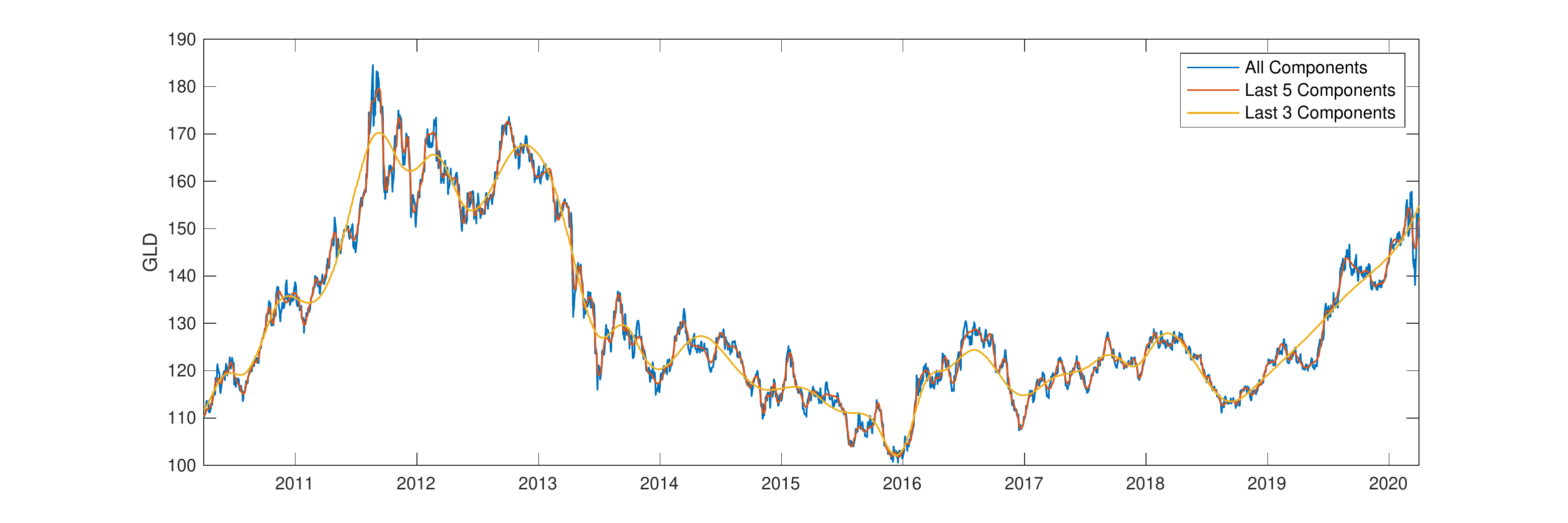}  \\
        \includegraphics[trim = 0.65cm   0.5cm    1.38cm  0.5cm, clip, width=5.85in]{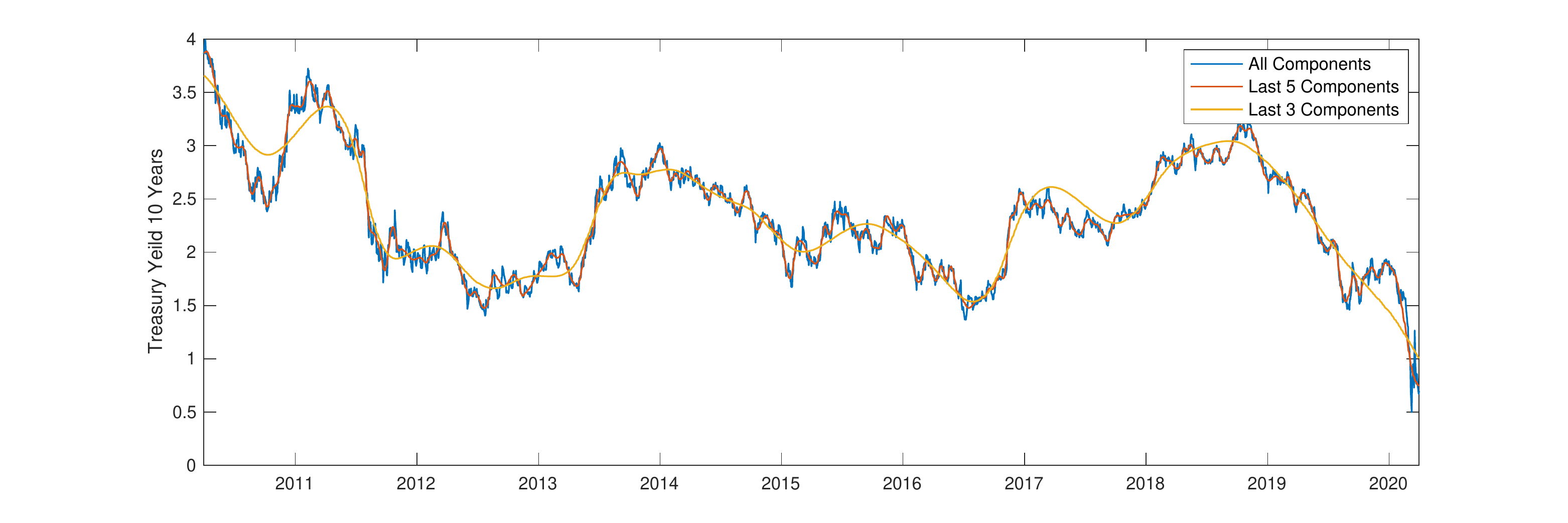} 
\caption{\small{Reconstruction of financial time series: S\&P 500, VIX, GLD and 10-year treasury yield. For S\&P 500 and GLD, the decomposition was implemented on the log price of the indices. The low-pass filtered time series by \eqref{lowpass} were then taken exponential to show the reconstruction.}}
    \label{fig_recon}
\end{figure}

In Fig. \ref{fig_recon}, we illustrate the low-pass filtered reconstruction of the four time series. This involves applying \eqref{lowpass} using different collections of components. Specifically, we have used the last 3 components and last 5 components including the residual terms.  We can see that, with more components included  in the reconstruction, the resulting time series resembles the original time series on a finer  timescale.  Compared to some other time series smoothing techniques, such as moving average, the CEEMD-based filter achieves smoothing without lags. Furthermore, CEEMD is capable of reconstructing time series with both trending or mean-reverting dynamics as shown in Fig. \ref{fig_recon}.  

\begin{figure}[ht]
\centering    %\includegraphics[trim = 1.1cm   2.4cm  1.5cm  1.5cm, clip, width=3.48in]{eemdSP500.eps}\\
    \includegraphics[trim = 1cm   1.8cm  1.5cm  1.5cm, clip, width=3.3in]{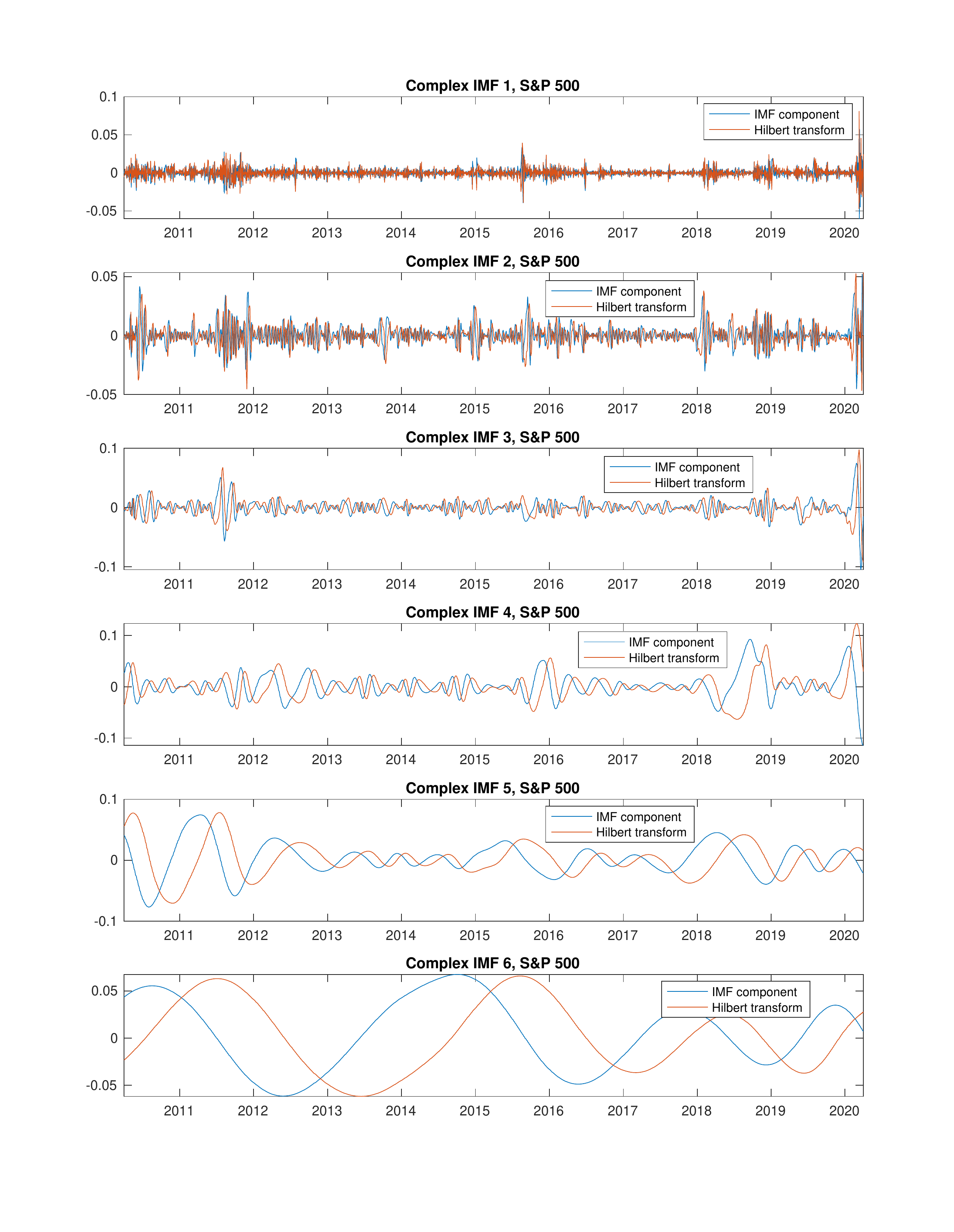}
    \includegraphics[trim = 1cm   1.8cm  1.5cm  1.5cm, clip, width=3.3in]{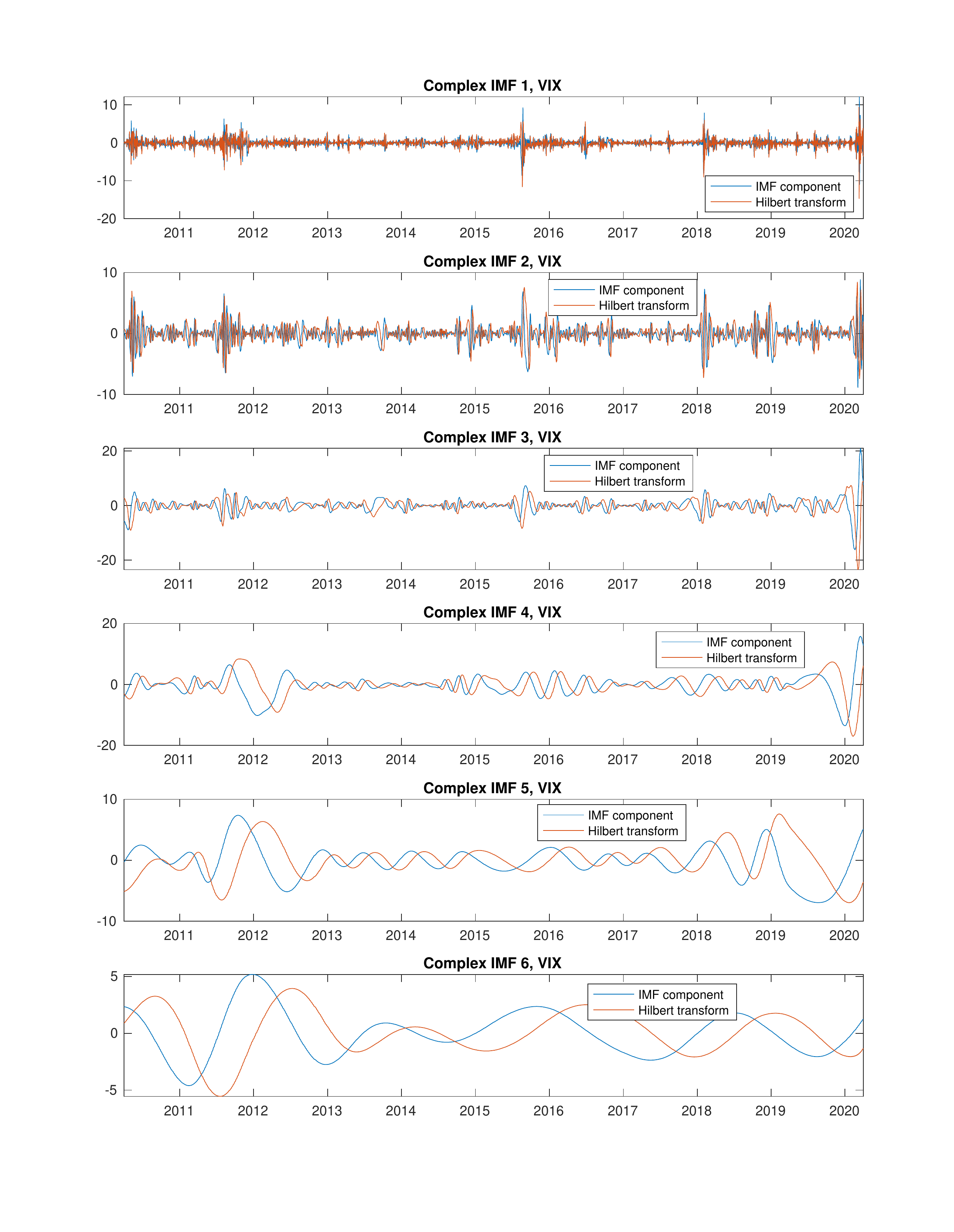}\\
     \includegraphics[trim = 1cm   1.8cm  1.5cm  1.5cm, clip, width=3.3in]{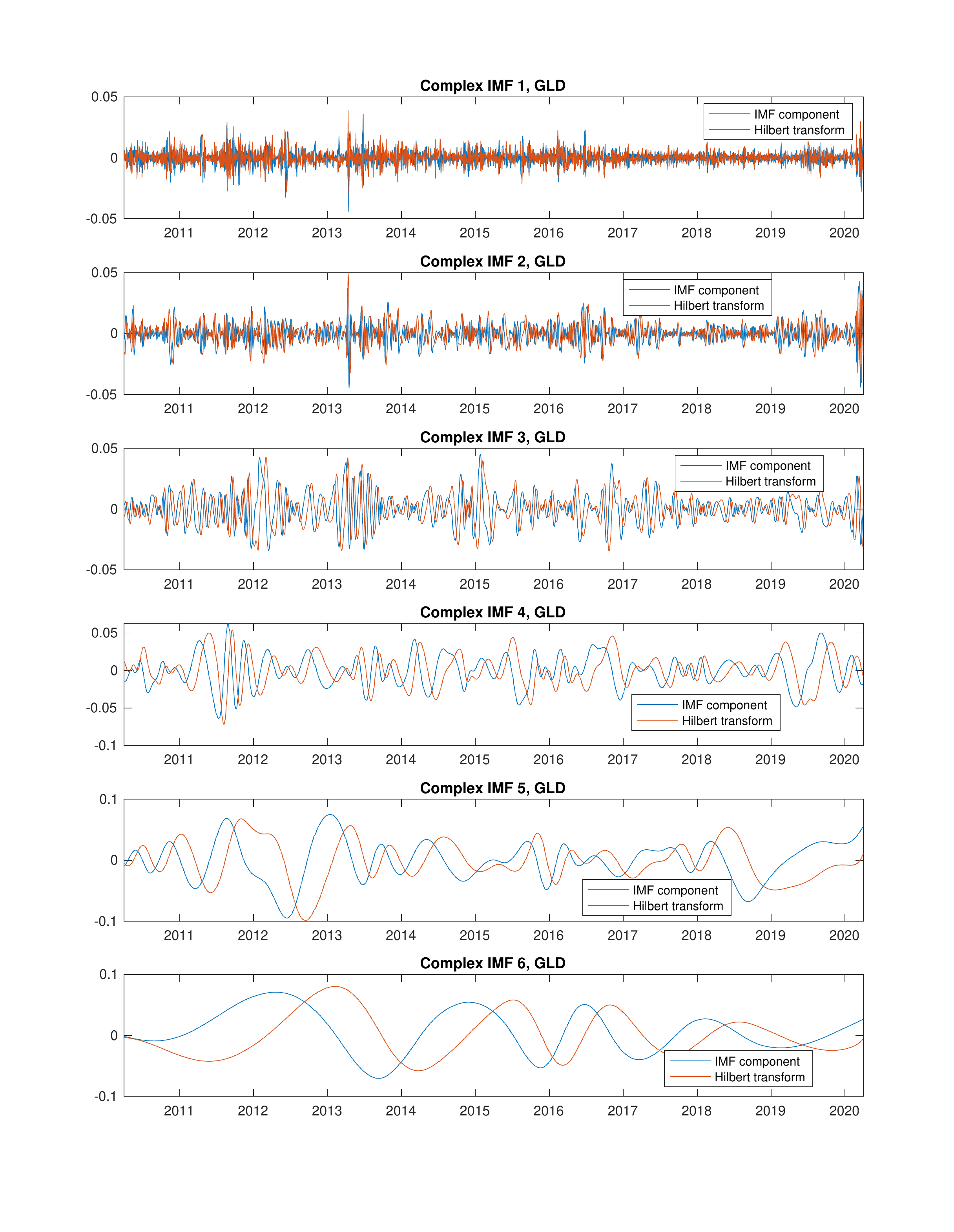}
     \includegraphics[trim = 1cm   1.8cm  1.5cm  1.5cm, clip, width=3.3in]{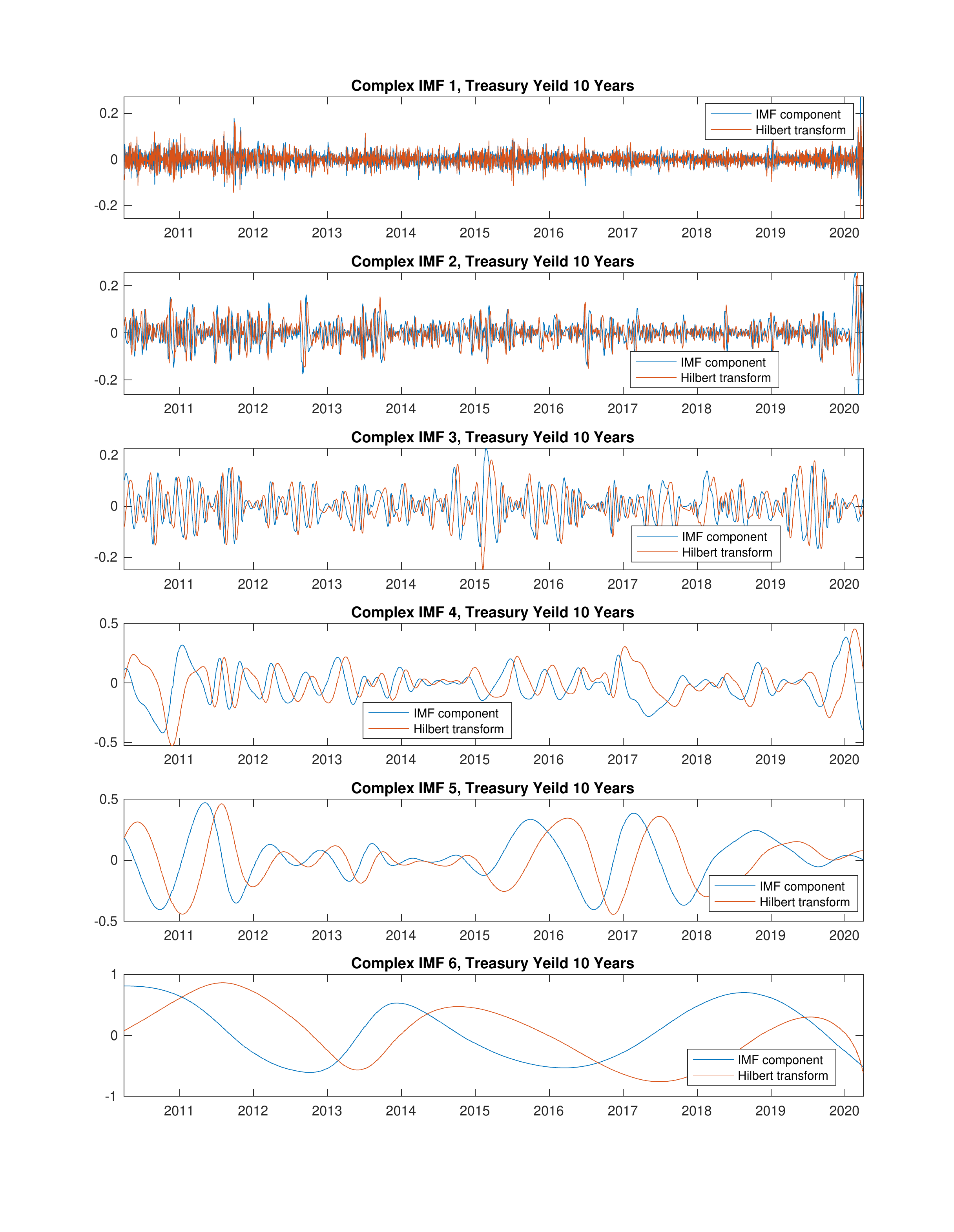}
    \caption{\small{Complex IMFs derived from formula \eqref{hilbert}, S\&P 500   (top left), VIX (top right), GLD   (bottom left), and 10-year treasury yield (bottom right). The real parts (IMF components) are plotted together with their corresponding imaginary parts (Hilbert transforms of IMFs).}}\label{fig_comp}
\end{figure}

Using formula \eqref{hilbert}, we implement Hilbert transform to the IMFs extracted from CEEMD. Following \eqref{eq_complex_imf}, we take the IMF components as the real parts and the corresponding Hilbert transform as the imaginary parts, and plot them together in Fig. \ref{fig_comp} as the complex IMFs which are analytic signals. We see that each real part (IMF component) and the corresponding imaginary part (Hilbert transform) are oscillations coupled together, where negative (positive) Hilbert transform corresponds to a increasing (decreasing) IMF component. The coupling behavior indicates the complex IMFs can be useful for time series forecasting.
	
\begin{figure}[ht]\centering
    \includegraphics[trim = 1cm   1.8cm  1.5cm  1.5cm, clip, width=3.3in]{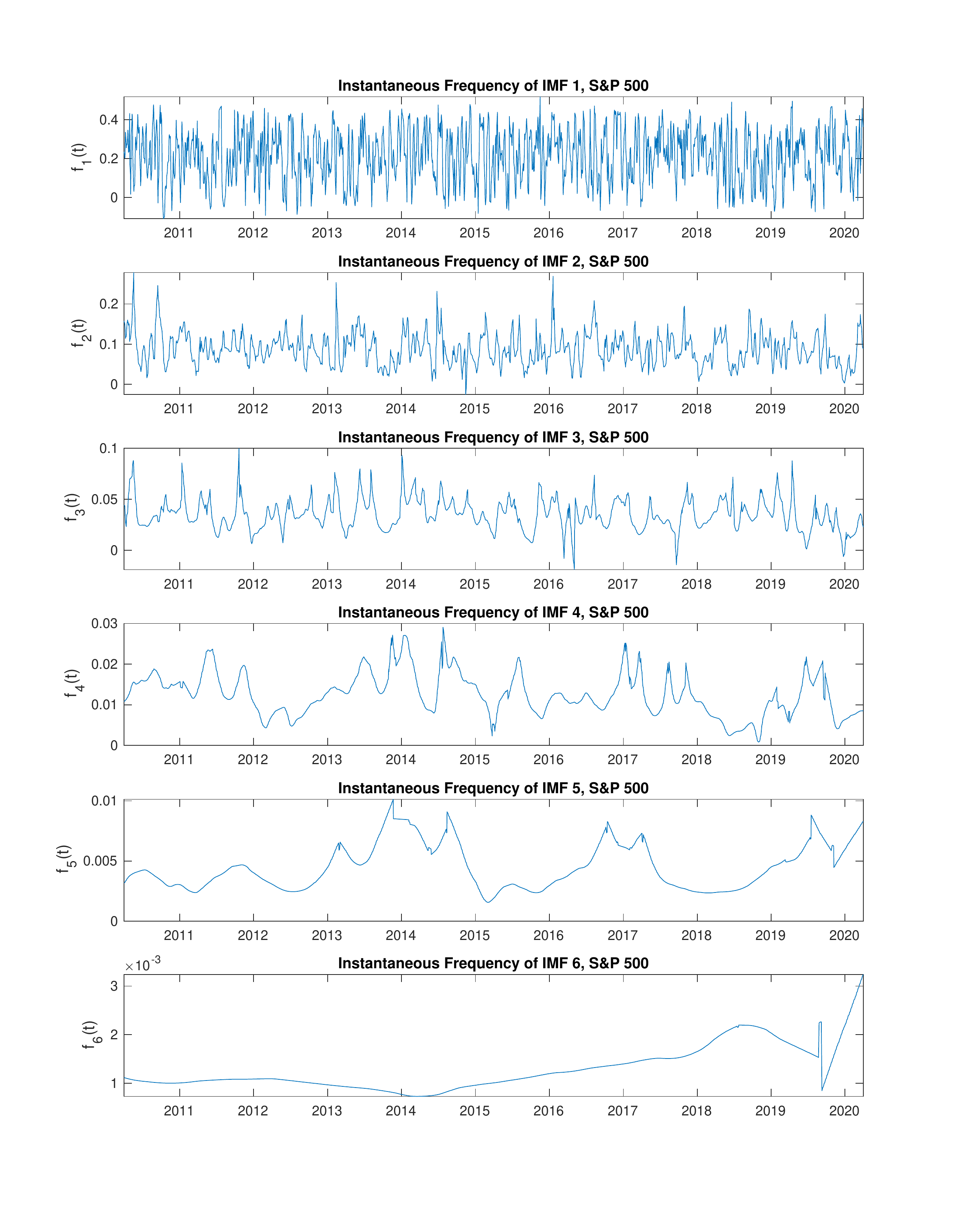}
    \includegraphics[trim = 1cm   1.8cm  1.5cm  1.5cm, clip, width=3.3in]{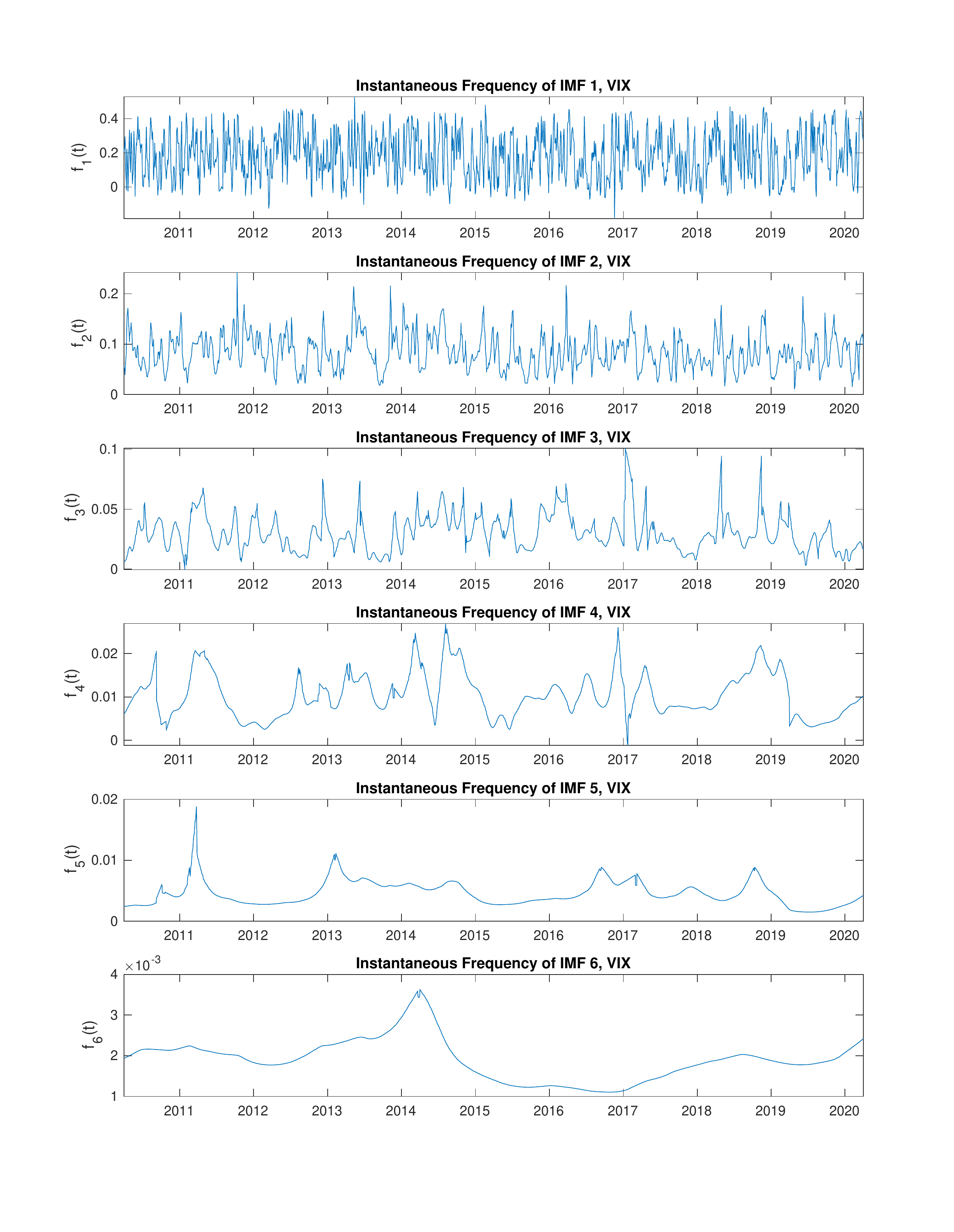}\\
     \includegraphics[trim = 1cm   1.8cm  1.5cm  1.5cm, clip, width=3.3in]{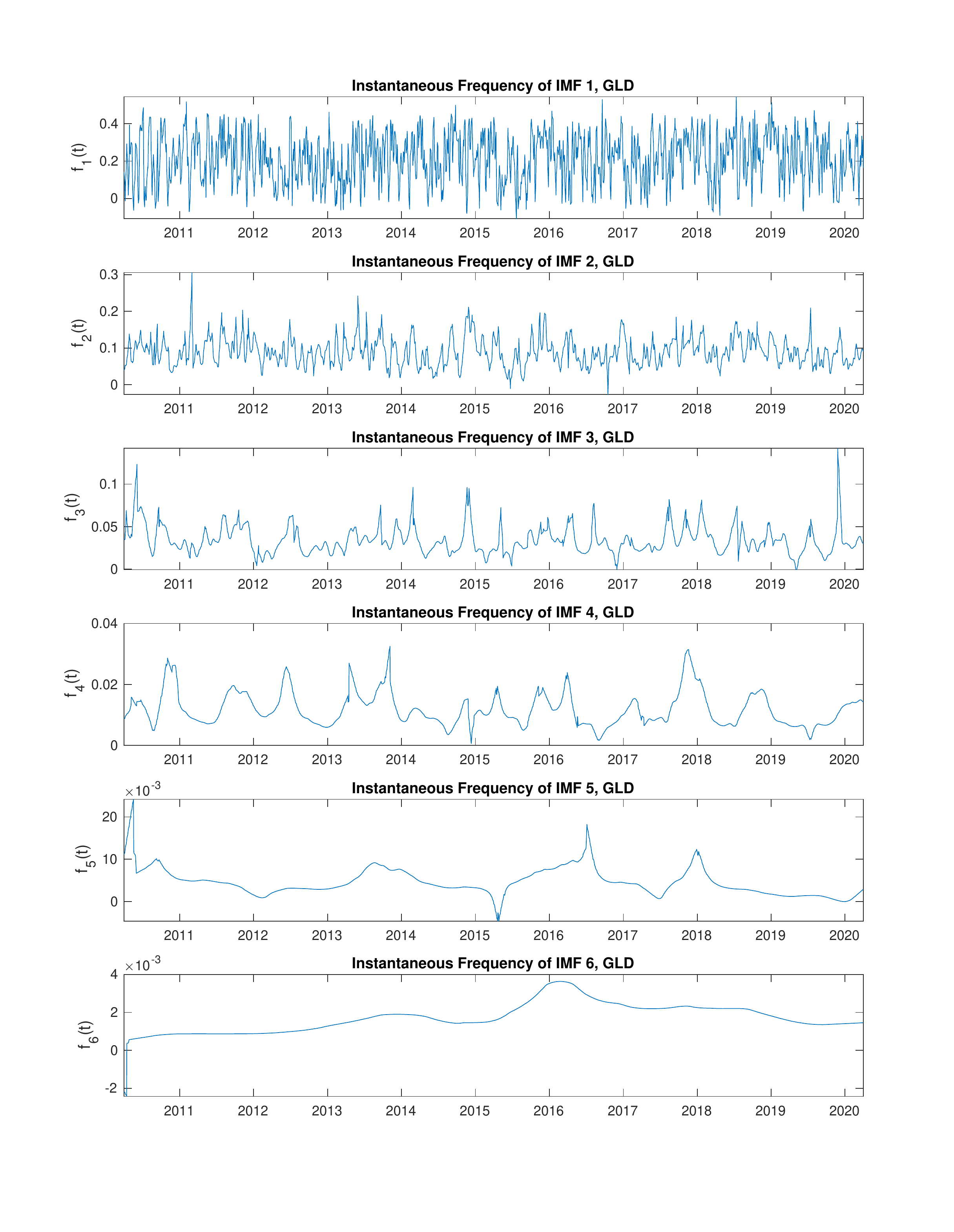}
     \includegraphics[trim = 1cm   1.8cm  1.5cm  1.5cm, clip, width=3.3in]{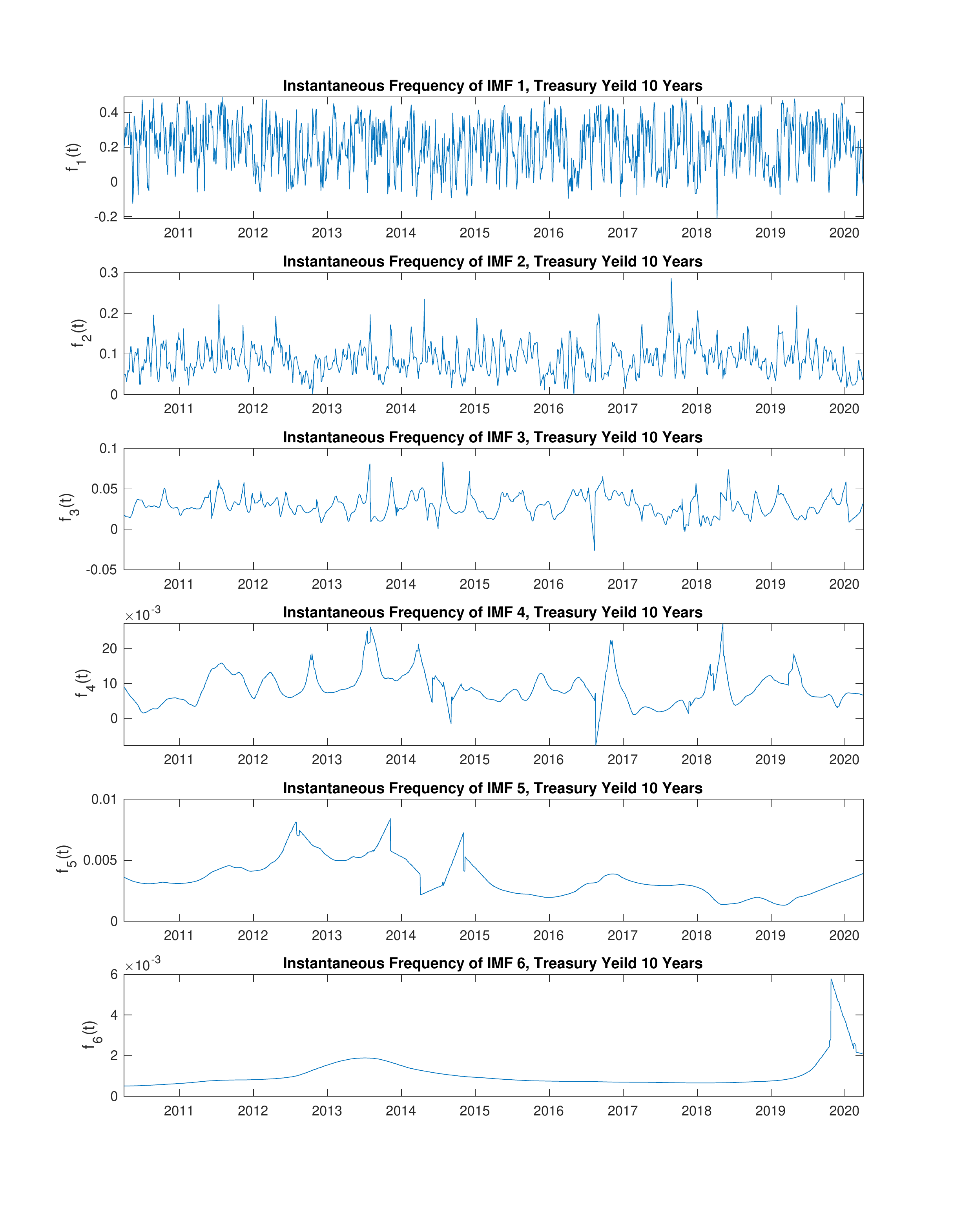}
    \caption{\small{The instantaneous  frequency associated with each IMF component derived from formula \eqref{freqEq} of S\&P 500, VIX, GLD, and TNX. From top to bottom, the instantaneous frequency correspond to modes with lower frequencies.}}\label{fig_hht_freq}
\end{figure}

\begin{figure}[t]
\centering
    \includegraphics[trim = 1cm   1.8cm  1.5cm  1.5cm, clip, width=3.3in]{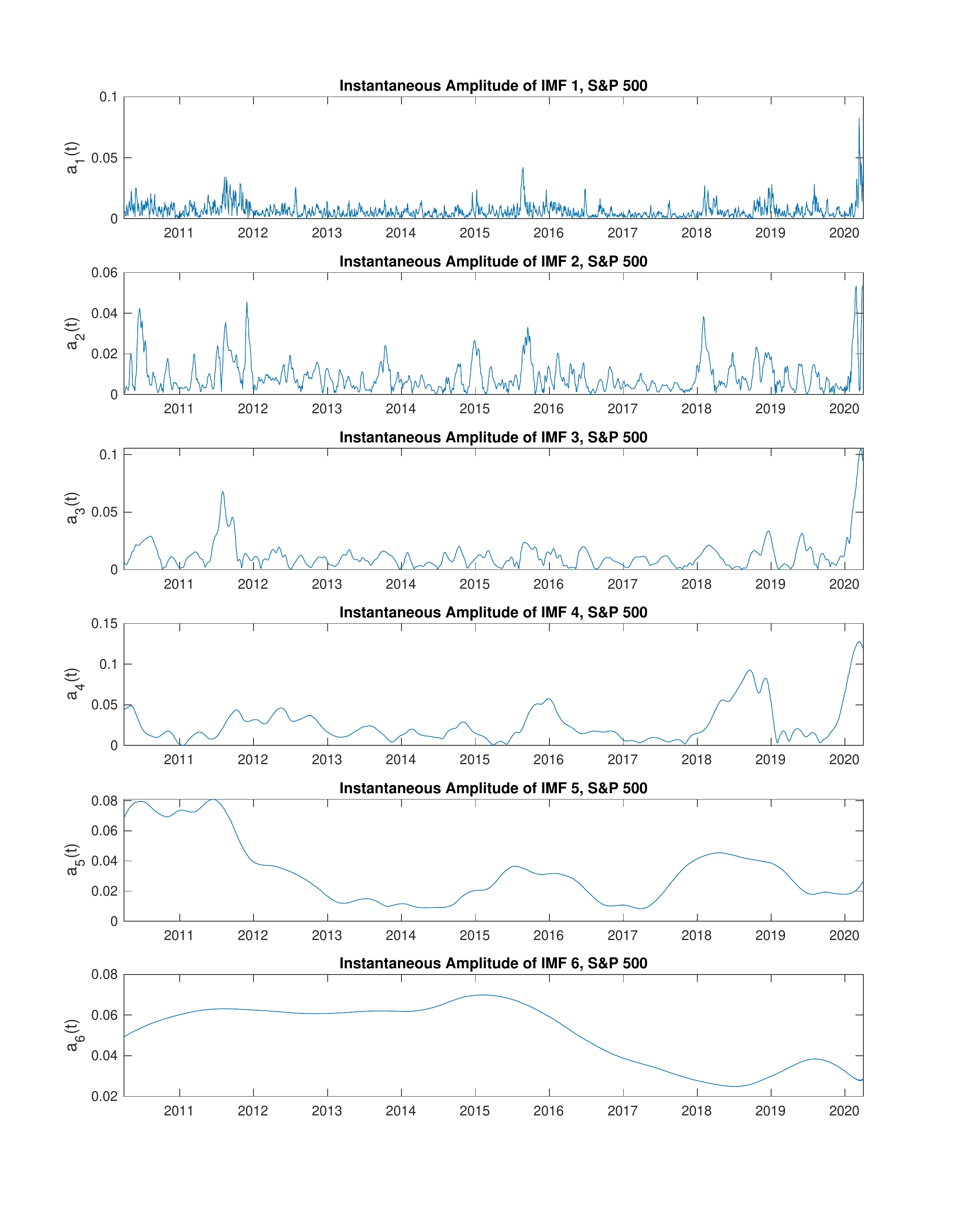}
    \includegraphics[trim = 1cm   1.8cm  1.5cm  1.5cm, clip, width=3.3in]{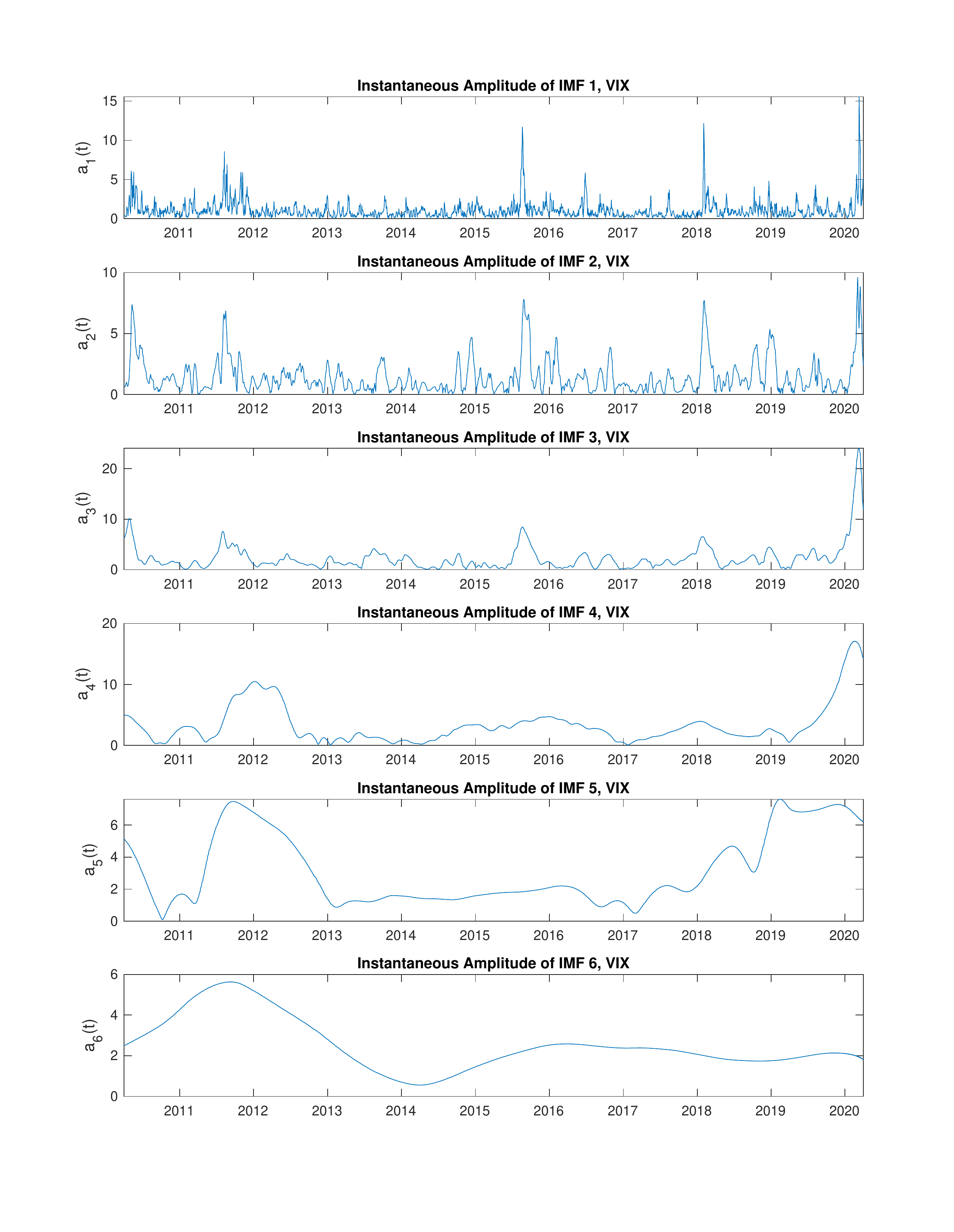}\\
     \includegraphics[trim = 1cm   1.8cm  1.5cm  1.5cm, clip, width=3.3in]{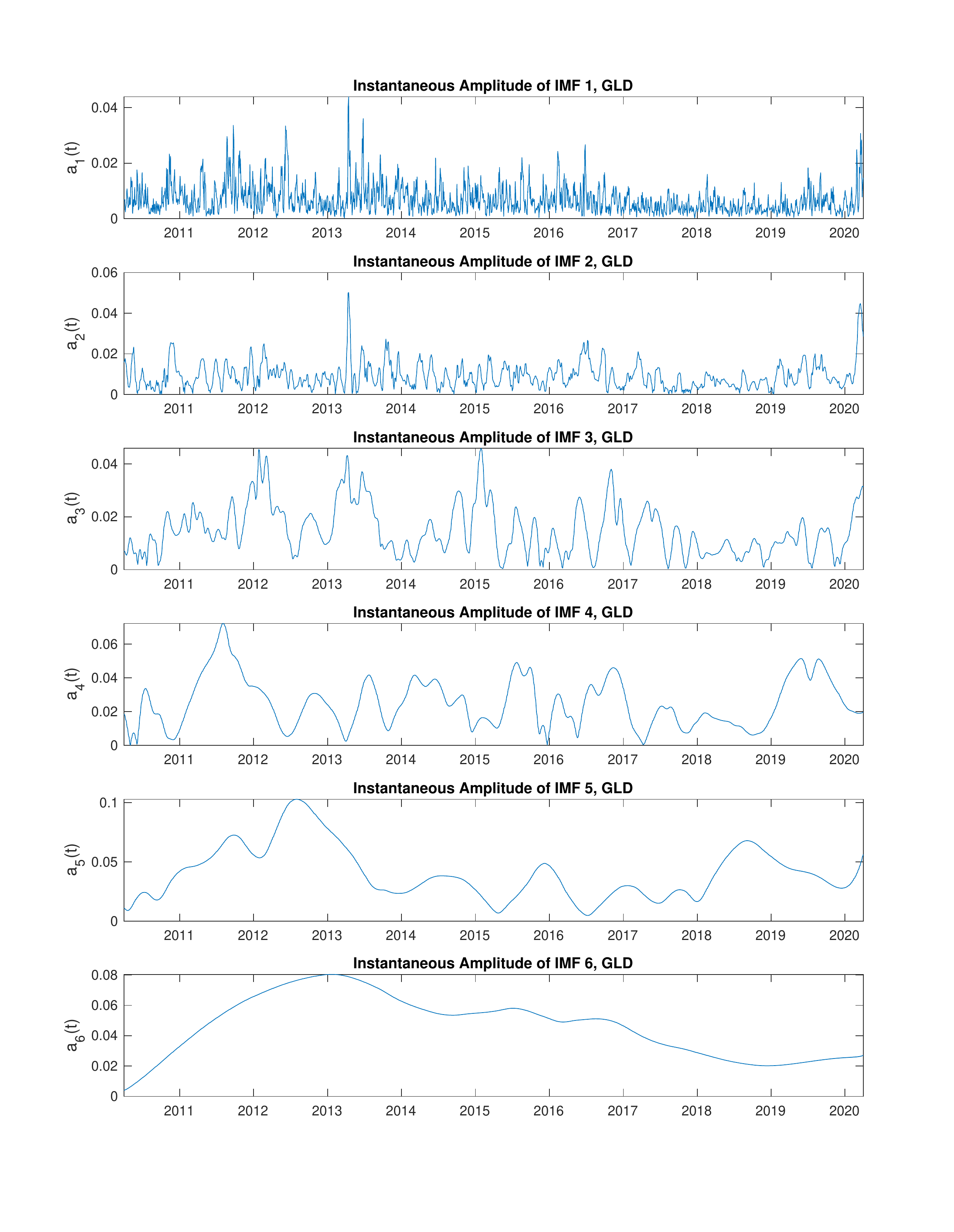}
     \includegraphics[trim = 1cm   1.8cm  1.5cm  1.5cm, clip, width=3.3in]{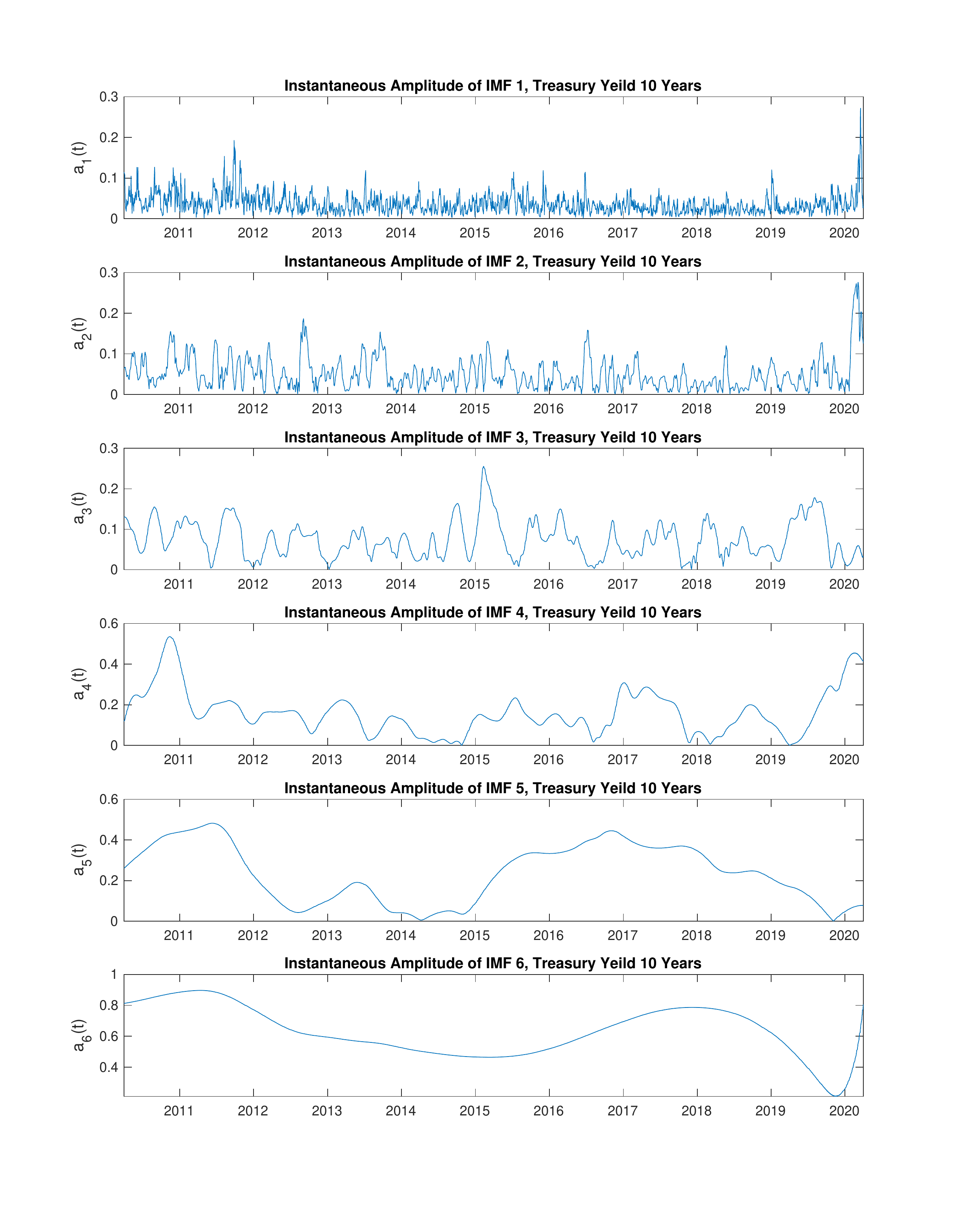}
    \caption{\small{The instantaneous  amplitude associated with each IMF component derived from formula \eqref{ampEq} of S\&P 500, VIX, GLD, and TNX. From top to bottom, the instantaneous amplitude correspond to modes with lower frequencies.}}\label{fig_hht_amp}
\end{figure}

The instantaneous  frequency $f_j(t)$ from \eqref{freqEq} associated with each IMF component $c_j(t)$ of the four financial time series are presented in Fig. \ref{fig_hht_freq}. From the 1st to 6th mode (top to bottom),   the  fluctuations of instantaneous  frequency tend to decrease, and the overall levels of instantaneous frequency are reduced.   Fig. \ref{fig_hht_amp} shows the instantaneous  amplitude $a_j(t)$ from \eqref{ampEq} associated with each IMF component $c_j(t)$ of the four financial time series. From the 1st to 6th mode (top to bottom),   the  fluctuations of instantaneous amplitude is reduced significantly. As such, the higher-order modes reflects a clearer long-term trend compared to lower-order modes. The fluctuations of the instantaneous amplitude and frequency are also less rapid than the IMF component itself, satisfying the slow modulation condition.
    
\subsection{End Effect of EMD}\label{sect-end}

EMD-based decomposition methods and HHT are adaptive for nonstationary systems. However, the algorithms are also   known for the issue of end effect, whereby   the decomposition and Hilbert transform have higher error when getting closer to the ends of the time series. This is because the sifting algorithm in Section \ref{sect-emd} uses interpolation in each iteration, requiring an anchor point beyond the time span. The end effect   can be relieved when the time series has enough length, and we are only interested in the properties in the interior. However, the error at the end can be a potential problem and may an additional challenge for forecasting.

To characterize the end effect, we need to consider the decomposition with the position in the time frame. Let's define the end effect factor as the position of point $t$ on the decomposition interval $[T_1, T_2]$:
\begin{equation}\label{eq-lambda}
\lambda(t) = \frac{2t - (T_1 + T_2)}{T_2 - T_1}.
\end{equation}
Note that $\lambda = 0$ at the middle point of the time frame, and $\lambda = \pm 1$ at the ends. Suppose the time series has the true IMF mode decomposition $x(t) = \sum_{j=1}^n c^*_j(t) + r^*_j(t)$, then the decomposition found by the CEEMD will be:

\begin{equation}\label{eq-end}
(c_1(t), \cdots, c_n(t),r_n(t)) \sim \mathcal{D}(c^*_1(t), \cdots, c^*_n(t),r^*_n(t), \lambda(t)),
\end{equation}
 which is a draw from the distribution due to the random nature of the noise-assisted algorithm. Denote the shorthand notice $\mathbf{c}(t) = (c_1(t), \cdots, c_n(t),r_n(t))$.  The error vector of $\mathbf{c}(t)$ is:
 \begin{equation}
\mathcal{E}(\mathbf{c}^*(t), \lambda(t)) = \mathcal{D}(\mathbf{c}^*(t), \lambda(t)) - \mathbf{c}^*(t).
\end{equation}
Note that $\mathbf{1}^T  \mathcal{E}(\mathbf{c}^*(t), \lambda(t)) = 0$ at any $t$, due to the exact decomposition in \eqref{eq-ceemd}, and $Var(\mathcal{E}_j(\mathbf{c}^*, \lambda))$ increase with $|\lambda|$ due to the end effect. 
 \clearpage

\subsection{Empirical Energy-Frequency Spectrum}\label{sect-sepctrum}
\begin{figure}[th]
    \centering
    \includegraphics[trim = 0.2cm   0cm  0.5cm  0.5cm, clip, width=3.1in]{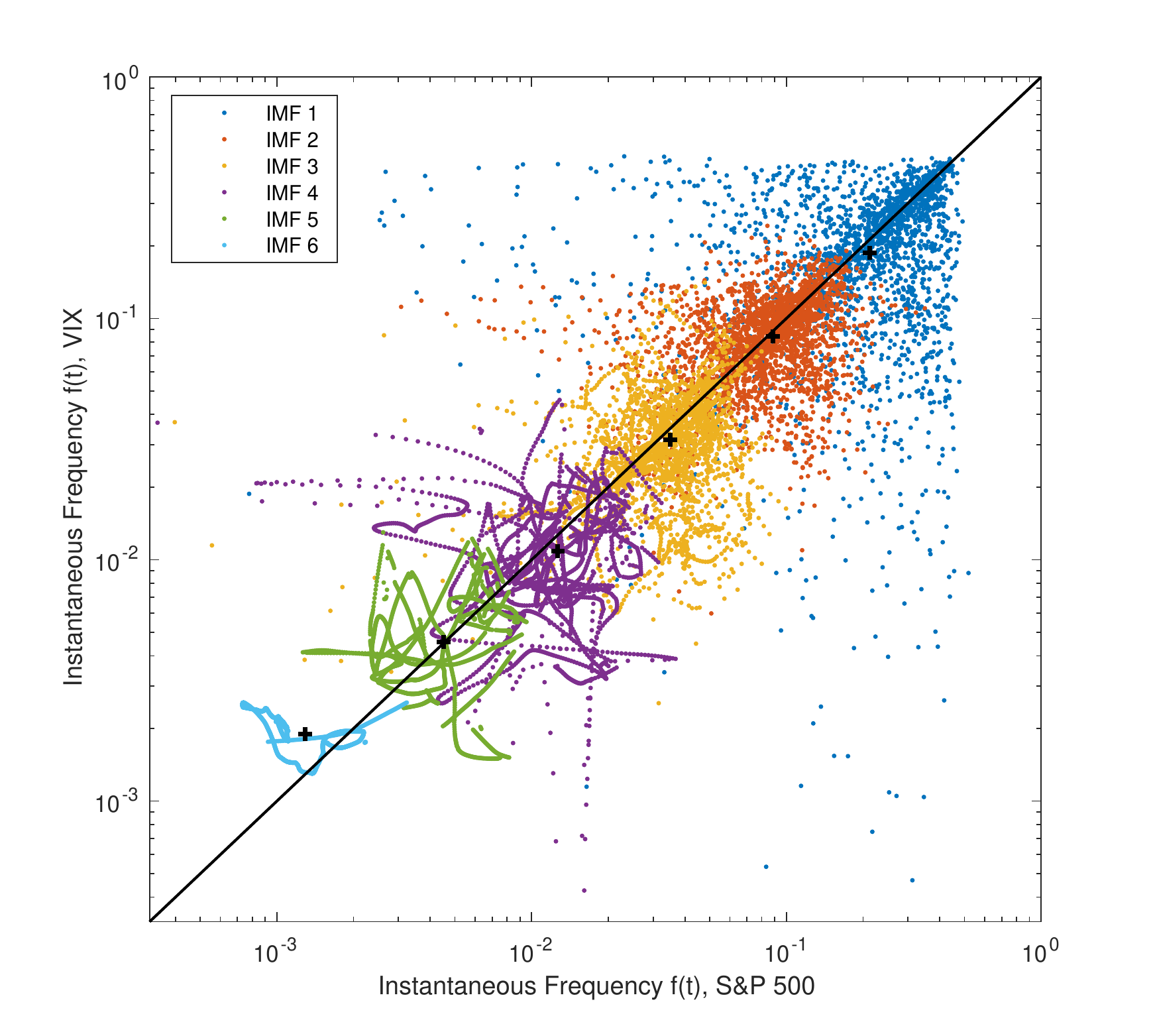}
    \includegraphics[trim = 0.2cm   0cm  0.5cm  0.5cm, clip, width=3.1in]{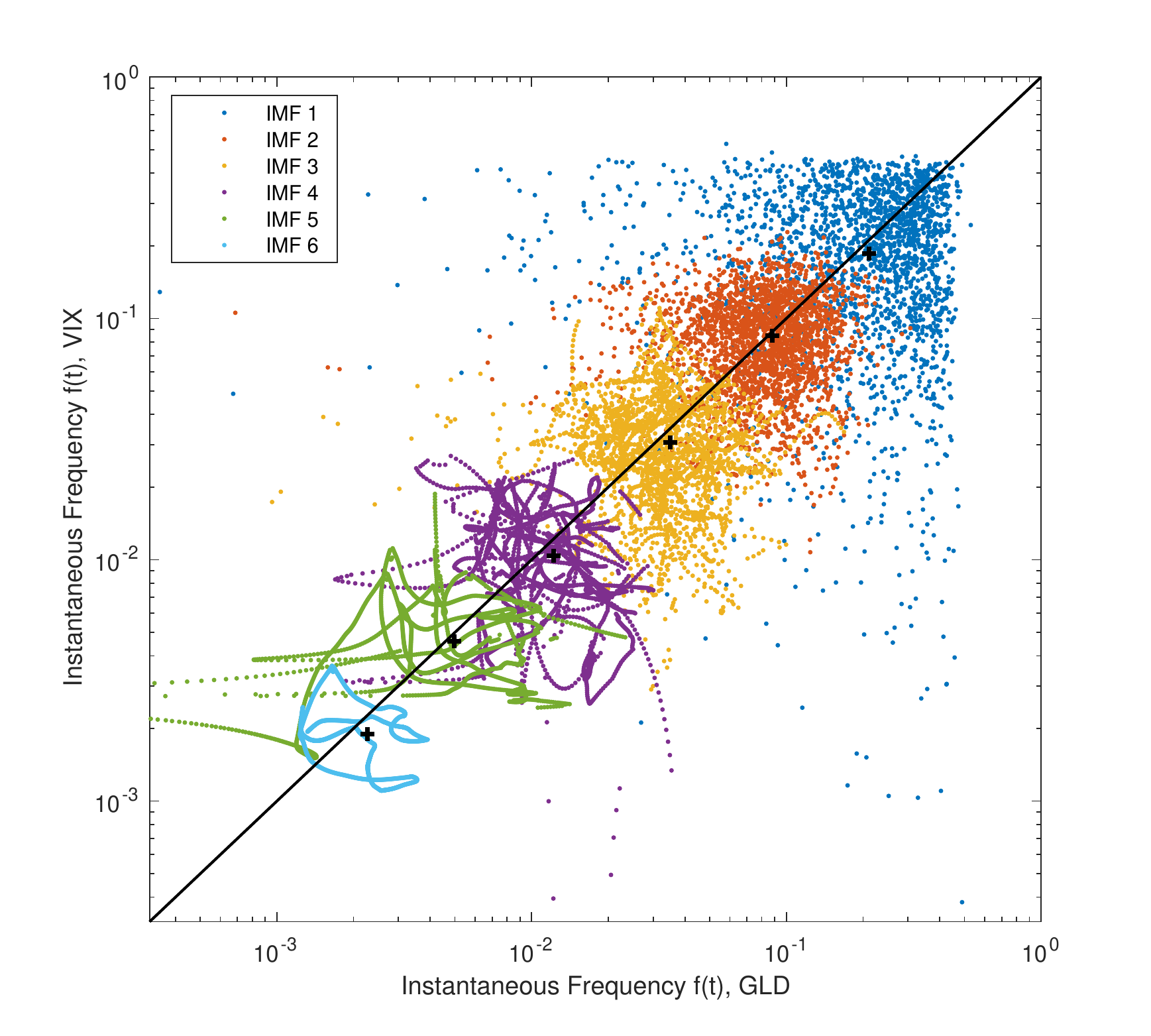}
        \caption{\small{Instantaneous frequency: S\&P 500  vs VIX (left) and GLD vs VIX (right).}}
    \label{fig:mode_spec2}
\end{figure}

For each financial time series, there is an instantaneous frequency profile. This allows us to compare the frequencies between any pair of indices, such as  S\&P 500 vs VIX  in Fig. \ref{fig:mode_spec2}. Each point on the plot is the pair  of instantaneous frequencies of S\&P 500 and VIX recorded at the same time. For each mode, there is one cluster of points (instantaneous frequencies) and the black cross marks the mean value of each cluster. These marks are located extremely close to the solid straight line of unit slope, which is a reference line for identical frequencies. This means that, while S\&P 500 and VIX have drastically different path behaviors, their dynamics share very similar frequencies on average across both long and short timescales. Similar observations are found for GLD when compared against VIX. 

% \begin{figure}[t]
%  \centering
%   \includegraphics[trim = 1.3cm   2.2cm  1.9cm  1.9cm, clip, width=3.45in]{ampSP500.eps}\\
% \caption{Instantaneous Amplitude}
% \label{fig_hht_amp}
% \end{figure}
 
%It worth note that the instantaneous frequency in the plots exhibit negative values occasionally, which is not physically meaningful. The reason is that the arc-tangent function in equation \ref{phaseEq} is not single-valued, which can take value on $\theta(t), \theta(t) \pm \pi, \theta \pm 2\pi, \cdots$. Even though we can interpreting $\theta(t)$ such that the change from $t$ to $t+1$ is as small as to be "continuous", there can be backward jumps. The instantaneous frequency $\omega(t)$, as the derivative of $\theta(t)$, is sensitive to the jumps and can therefore be negative. To reduce the error in instantaneous frequency computation, we use the robust locally weighted scatterplot smoothing (robust LOWESS) introduced by \cite{cleveland1979robust}. The smoothed instantaneous frequency is used in the analysis starting this point.

%Hilbert-Huang transform of the modes of S\&P500 log price, from January the 4th, 2010 to December the 31st, 2019. For each mode, we show the imaginary conjugate part of the EMD modes, instantaneous frequency and amplitude estimation from Hilbert transform.
 %Plot (a) is each figure shows the Hilbert transform by using formula \ref{hilbert}. The EMD modes is treated as $X(t)$ in equation \ref{hilbert}, and the imaginary part is the $Y(t)$ in equation \ref{hilbert} from Hilbert transform. We plot them together to show the coupled oscillations.

\begin{figure} [t]
 \centering
    \includegraphics[trim = 0.5cm   0.5cm   1.3cm  0.5cm, clip,width=3.1in]{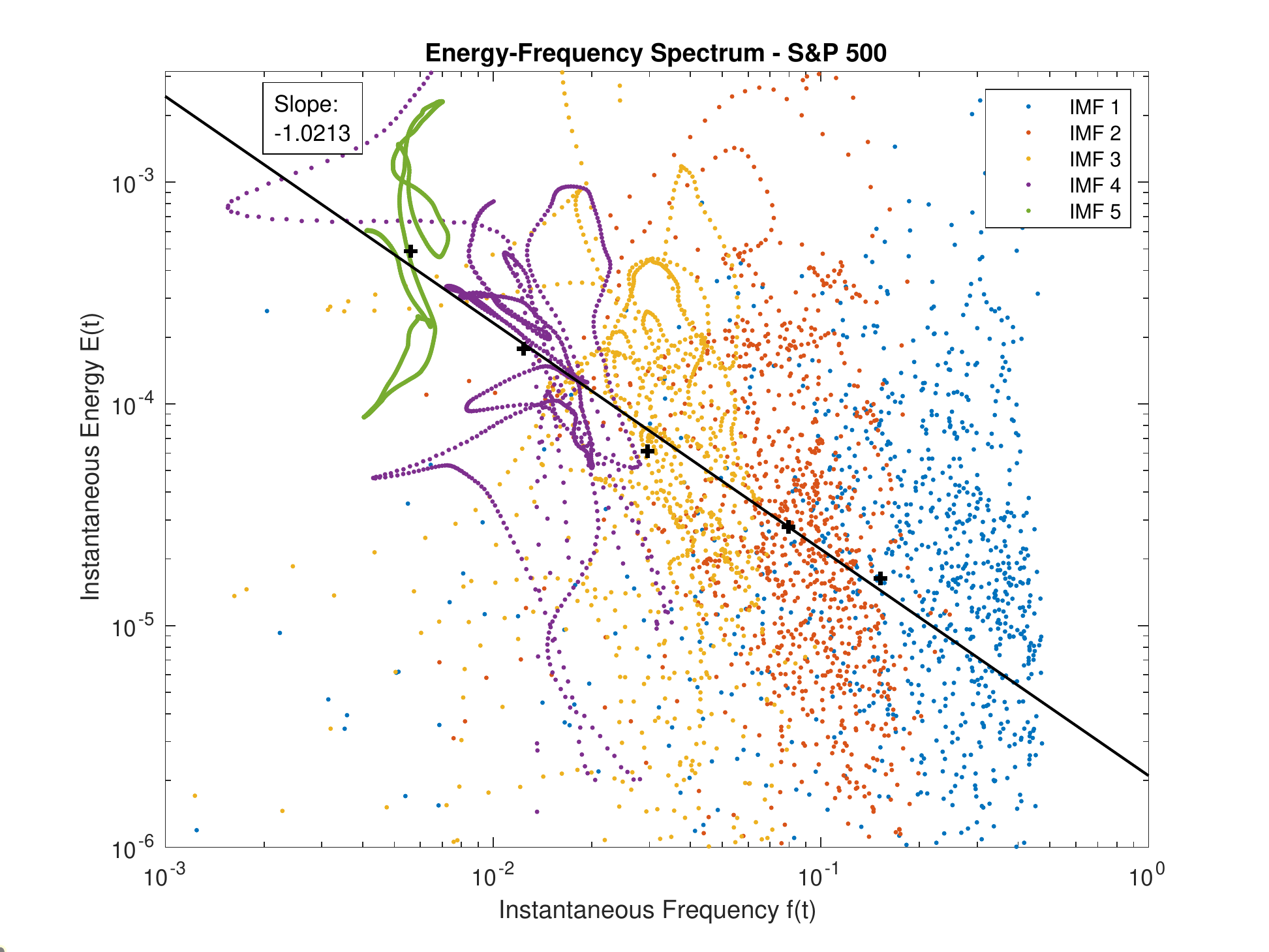}
        \includegraphics[trim = 0.5cm   0.5cm   1.3cm  0.5cm, clip,width=3.1in]{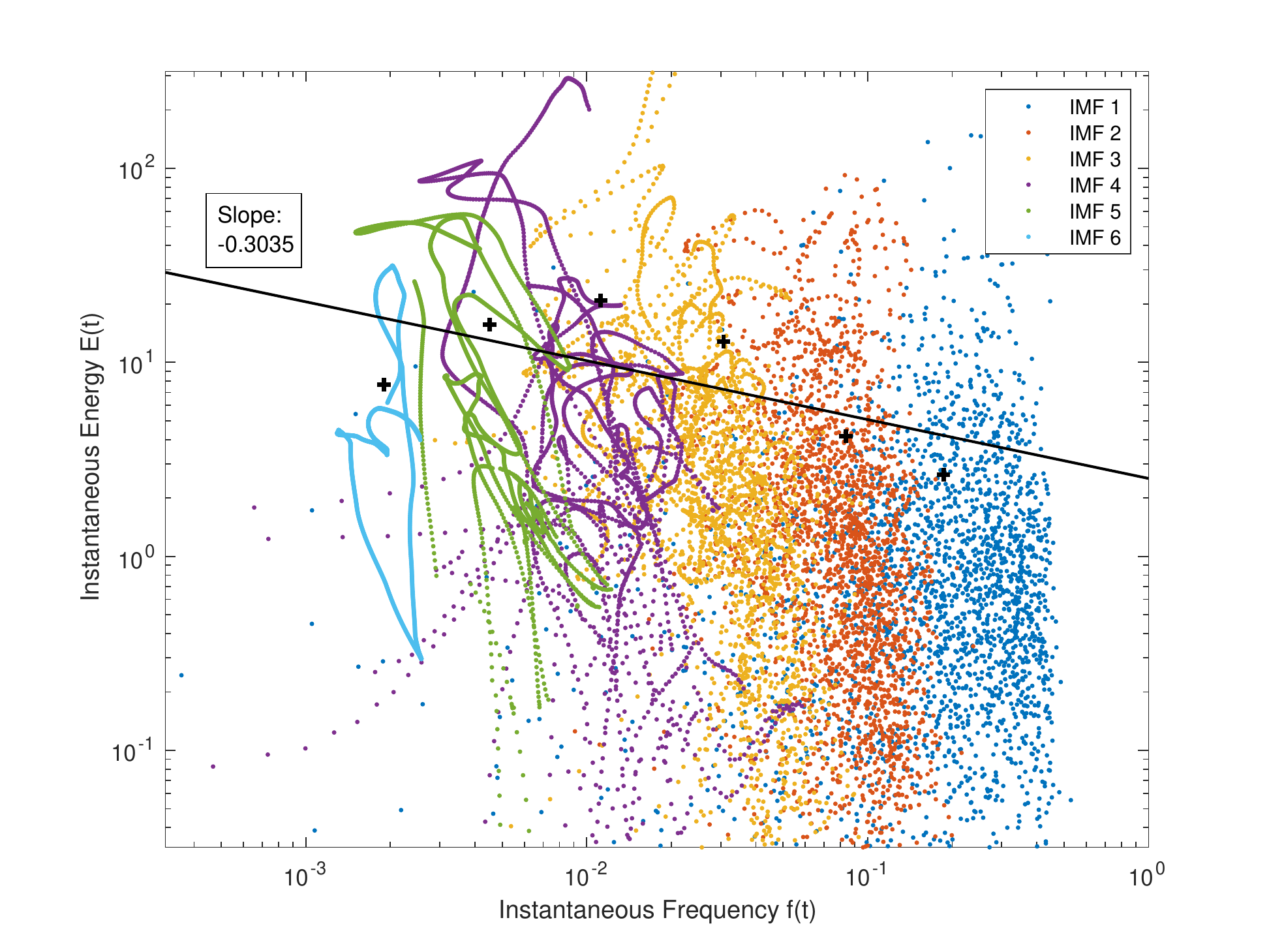}\\
        \includegraphics[trim = 0.5cm   0.5cm   1.3cm  0.5cm, clip,width=3.1in]{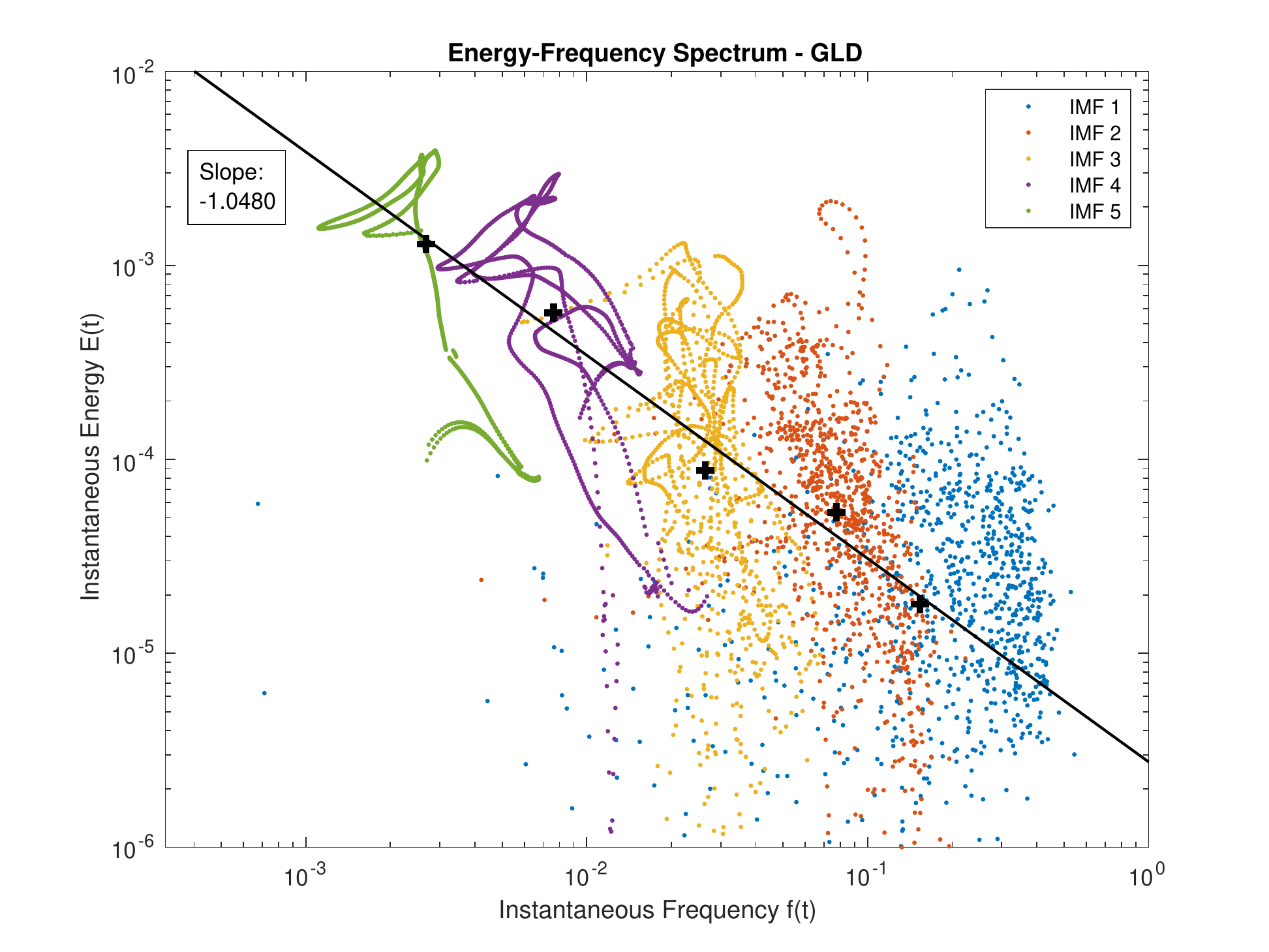}
        \includegraphics[trim = 0.5cm   0.5cm   1.3cm  0.5cm,clip,width=3.1in]{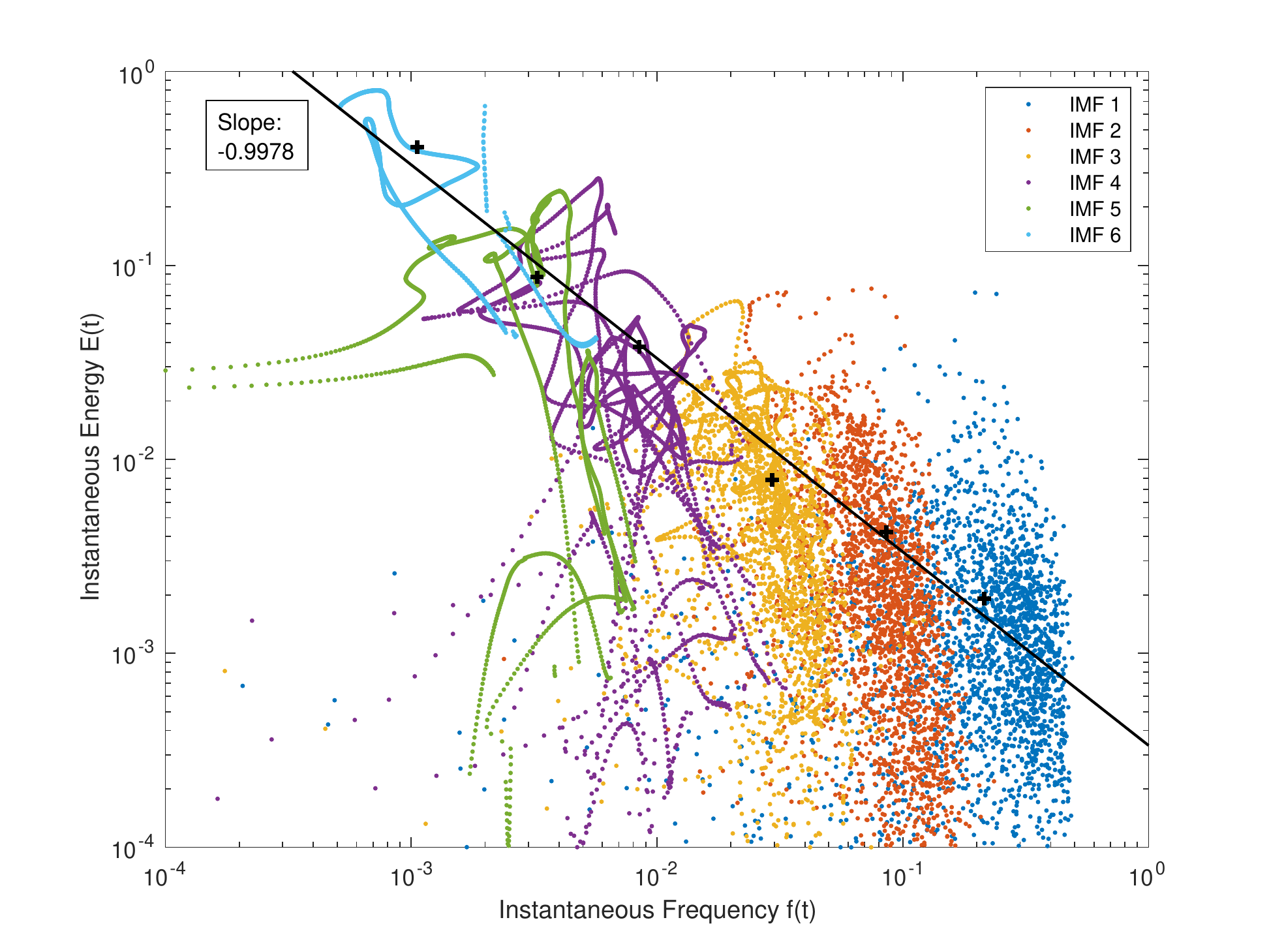}\
        \caption{\small{Instantaneous energy-frequency spectrum for S\&P 500 (top left), VIX (top right), GLD (bottom left), and 10-year treasury yield (bottom right). The black cross marks the mean of the cluster of points for each mode.}}
    \label{fig:mode_spec}
\end{figure}

Next, we show in Fig. \ref{fig:mode_spec}   the instantaneous energy-frequency spectrum of the S\&P 500 (top left), VIX (top right), GLD (bottom left), and 10-year treasury yield (bottom right) across all IMF components. Each point on the plots is a pair of $(f_j(t), E_j(t))$, for  mode $j=1,\cdots,n$, and time $t=1,\cdots,T$. For each mode, the black cross marks the mean of each cluster of points. The clusters from the right to left correspond to  higher modes.  

For S\&P 500, from IMF 1 to 6, the average instantaneous frequency decreases from  $2.122\times 10^{-1}$ to $1.287\times 10^{-3}$ respectively, indicating cycles on timescales from 5 trading days (weekly behavior) to over 750 trading days (about 3 years). Since the decomposition is implemented over a 10-year period, the residual term reflects the long-term trend, possibly driven by economic cycles. The solid line obtained from linear regression has a negative slope for all the four financial time series. This means that the instantaneous energy $E(t)$ tends to decrease as the instantaneous frequency $f(t)$ increases.

Comparing the four indices, S\&P 500, GLD and 10-year treasury rate exhibit a more rapid dissipation of energy with respect to frequency than VIX. This reveals that the rapid fluctuations in S\&P 500, GLD and 10-year treasury rate tend to have much smaller magnitude than those for VIX. Moreover, S\&P 500 has a significantly wider range of instantaneous energy than VIX. This is consistent with the fact that  S\&P500 has risen from 1100 to 3000 while VIX has stayed between 10 and 40 over the 10-year period. In terms of power spectrum, S\&P 500, GLD and 10-year treasury rate are closer to the famous $1/f$ power spectrum, while VIX shows significantly different behavior. More interestingly, the residue with respect to the linear regression on the log-log plot of VIX shows significant concavity. These difference all indicate the volatility process follows a very different mechanism than the price processes.

For higher modes (lower frequencies),  the instantaneous energy and frequency have smoother paths over time,  giving rise to more continuous trajectories on the left-side of the plot. This is consistent with the fact that the fluctuation of instantaneous amplitude and instantaneous frequency are reduced from lower to higher modes, as shown in Fig. \ref{fig_hht_amp} and \ref{fig_hht_freq}. Using the language of dynamical systems, we can view the energy-frequency spectra as movements on the phase plane $(f, E)$, which serves as an alternative representation of the dynamics. 
 
%\begin{figure}
%    \includegraphics[trim = 1cm   2.5cm   1cm  1.5cm, clip, width=3.45in]{compSP500.eps}
%\caption{Complex Conjugate EMD Modes. } 
%\label{fig_hht_cc}
%\end{figure}

\subsection{Comparison to Other Methods}
We end this section by highlighting the differences among the HHT-based method,  Fourier transform, and wavelet transform. As summarized in  Table \ref{table:compare}, HHT allows for nonlinearity, nonstationarity. The HHT bases are determined in adaptive form with respect to the time series, rather than being prespecified as in the Fourier and wavelet transforms. In addition, HHT produces a sparse spectrum with finitely many of modes, which is a useful property for our study of financial time series. The adaptiveness and sparsity also allow  for practical feature generation for time series forecasting, which is the main reason for our use of HHT in this study.
\begin{table}
\centering
\begin{small}
\begin{tabular}{r|cccc}
\hline
 &  Nonstationarity & Nonlinearity & Basis & Spectrum \\
\hline
\hline
Fourier & No & No & A priori & Global; Dense \\
\hline
Wavelet & Yes & No & A priori & Regional; Dense \\
\hline
HHT & Yes & Yes & Adaptive & Local; Sparse \\
\hline
\end{tabular}
\caption{Comparison between different transforms for time series.}
\label{table:compare}
\end{small}
\end{table}

\section{Machine Learning Using HHT Features}\label{sect-ml}
\begin{figure}[ht]
    \centering
    \includegraphics[width= 0.45\textwidth]{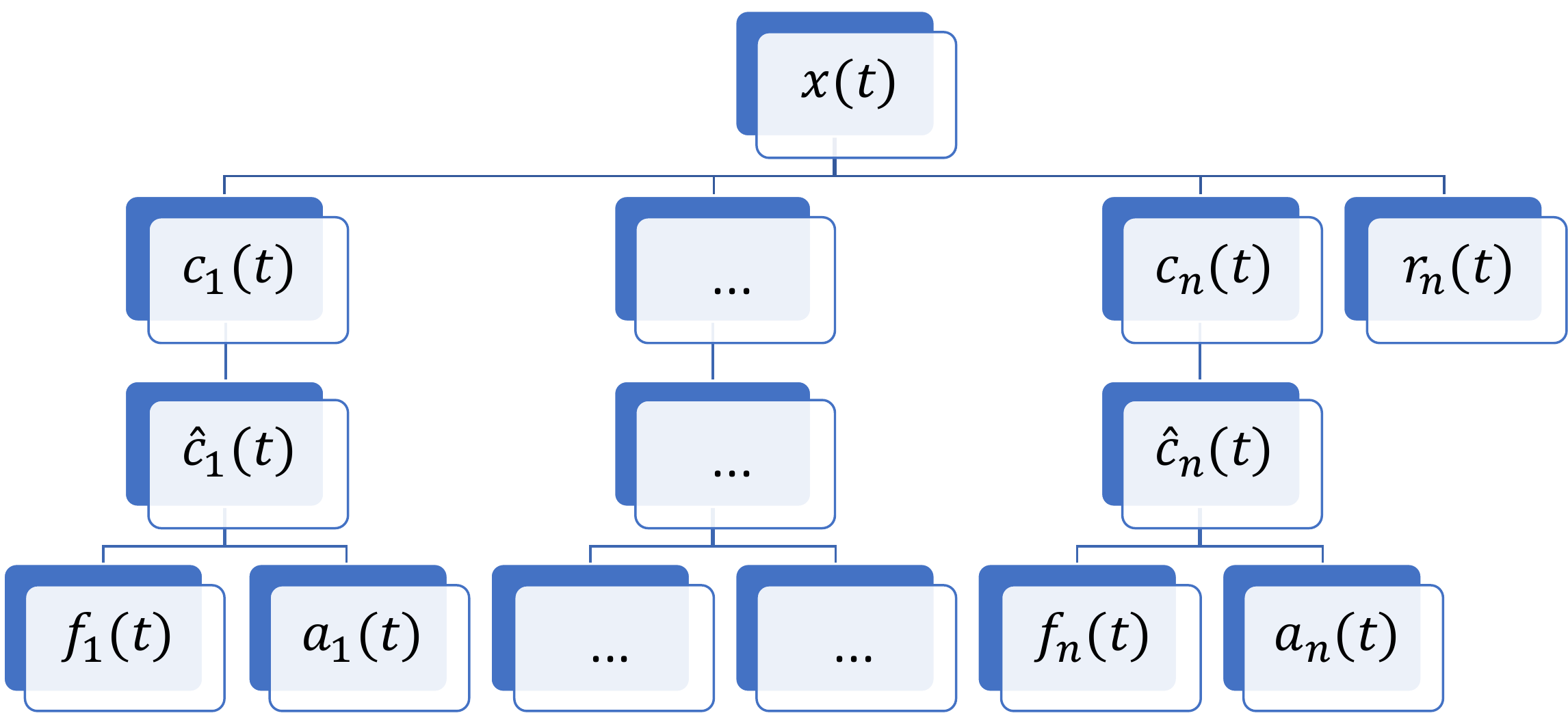}
    \caption{\small{The collection  of HHT features derived from the original time series $x(t)$.}}
    \label{HHTFeatures}
\end{figure}

The implementation of HHT-enhanced ML predictions involves two main steps. In the first step, HHT is applied to the original time series $x(t)$. CEEMD decomposes $x(t)$ into $n$ IMFs $c_1(t), \cdots, c_n(t)$ plus the residual term $r_n(t)$, and then  Hilbert transform on the IMFs yields the corresponding imaginary counterparts $\hat{c}_1(t), \cdots, \hat{c}_n(t)$, along with the instantaneous amplitude $a_1(t), \cdots, a_n(t)$, and  instantaneous frequency $f_1(t), \cdots, f_n(t)$. These outputs represent the \emph{HHT feature set} derived from the time series $x(t)$ (see Fig. \ref{HHTFeatures}). The following sections discuss the second step of training and testing machine learing models on HHT features.

\subsection{Training and Testing }
After the extracting of the HHT features,  a machine learning model is chosen to analyze the times series and make predictions for future time periods. Specifically, the goal is to predict the change in the next period $\Delta x(t+1) := x(t+1) - x(t)$, going forward in time $t$. Traditional time series prediction takes the values of $x$ in the past $\tau$ periods $x^{(t,\tau)}:= (x(t-\tau+1),\cdots,x(t))$ as input. In contrast,  our approach uses only the HHT features as predictors, rather than the original time series. To this end, we use the following shorthand notations:
\begin{align}
    c^{(t,\tau)}_j &:= (c_j(t-\tau+1),\cdots,c_j(t)), \\
    c^{(t,\tau)}_{(j,l)} &:= (c^{(t,\tau)}_j, \ldots, c^{(t,\tau)}_k),
%    \hat{c}^{(t,k)}_j &:= [\hat{c}_j(t-k+1),\cdots,\hat{c}_j(t)], \\
%    a^{(t,k)}_j &:= [a_j(t-k+1),\cdots,a_j(t)], \\
%    \omega^{(t,k)}_j &:= [\omega_j(t-k+1),\cdots,\omega_j(t)].
\end{align}
and likewise for other HHT features $\hat{c}, a$ and $f$. 

The objective is to minimize the expected $l_2$-loss of $\Delta x(t+1)$ prediction, given information up to time $t$. This leads to the optimization problem:
\begin{equation}\notag
    \min_{g\in \mathcal{G}} \mathbb{E} [(\Delta x(t+1)- g(c^{(t,\tau)}_{(1,n)},\hat{c}^{(t,\tau)}_{(1,n)}, a^{(t,\tau)}_{(1,n)},f^{(t,\tau)}_{(1,n)}))^2|\mathcal{F}_t],
\end{equation}
where $\mathcal{G}$ is the functional class specified by the ML model so that each $g$ is a prediction function determined within the training period. The filtration of information up to time $t$ is denoted by $\mathcal{F}_t$. For implementation with real data,  we do not know the distribution of input features and the target output, so the expectation is  estimated by the sample average. Specifically, we split the time series into a training section from $t=1$ to $t=T_1$, and a testing section after that from $t=T_1+1$ to $t=T_1+T_2$. The ML models learn on the training set by solving
\begin{equation}\notag
    \min_{g\in \mathcal{G}} \frac{1}{T_1}\sum_{t=1}^{T_1} (\Delta x(t+1)- g(c^{(t,\tau)}_{(1,n)},\hat{c}^{(t,\tau)}_{(1,n)}, a^{(t,\tau)}_{(1,n)},f^{(t,\tau)}_{(1,n)}))^2.
\end{equation}
The optimal predictive function found, $g_*$,  is then tested on the unseen data. The performance is measured in terms of mean squared error (MSE):
\begin{equation}\notag
     \frac{1}{T_2}\sum_{t=T_1+1}^{T_1+T_2} (\Delta x(t+1)- g_*(c^{(t,\tau)}_{(1,n)},\hat{c}^{(t,\tau)}_{(1,n)}, a^{(t,\tau)}_{(1,n)},f^{(t,\tau)}_{(1,n)}))^2.
\end{equation}

\subsection{Extrapolating Prediction}
\label{sect-extra1}
The training and testing framework discussed above is useful in validating the predictive power of HHT and perform feature selection. However, in real-time forecasting, we are only able to do the decomposition up to the current time $T$, and the testing is the one-shot extrapolation into the future $T+1$ where no decomposition is available. The above-mentioned procedure and most of the previous studies suffered from information leakage from the future that enhanced the prediction at the current time, due to the interpolation property described in Section \ref{sect-end}. 

As discussed above, the traditional train/test split does not work for time series prediction using HHT features. To test the extrapolation power of HHT features, we propose the following procedure of training and testing:

\begin{itemize}
\item For $t = T_1+1,\cdots, T_1+T_2$:
\begin{itemize}
\item Implement CEEMD and HHT on $x(s)$ for $s \in [t-T, t-1]$. 
\item Train machine learning model on $[t-T, t-1]$.
\begin{equation}\notag
    \min_{g\in \mathcal{G}} \frac{1}{T}\sum_{s=t-T}^{t-1} (\Delta x(s)- g(c^{(s-1,\tau)}_{(1,n)},\hat{c}^{(s-1,\tau)}_{(1,n)}, a^{(s-1,\tau)}_{(1,n)},f^{(s-1,\tau)}_{(1,n)}))^2.
\end{equation}
\item Test one-shot prediction of $\hat{x}(t) = g_*(c^{(t-1,\tau)}_{(1,n)},\hat{c}^{(t-1,\tau)}_{(1,n)}, a^{(t-1,\tau)}_{(1,n)},f^{(t-1,\tau)}_{(1,n)})$.
\end{itemize}
\item Evaluate testing mean square error $\frac{1}{T_2}\sum_{t=T_1+1}^{T_1+T_2} (x(t) - \hat{x}(t))^2$. 
\end{itemize}

\subsection{Machine Learning Models}
Our approach is generally applicable to all ML models that take multidimensional features as inputs. To illustrate our methodology, we implement several ML models for time series predictions. Specifically, we tested three different types of nonlinear regression models, namely the regression tree ensemble (RTE), the support vector machine (SVM) regression, and the long short-term memory (LSTM) neural network. 

\begin{figure}
    \centering
    \includegraphics[width=0.7\textwidth]{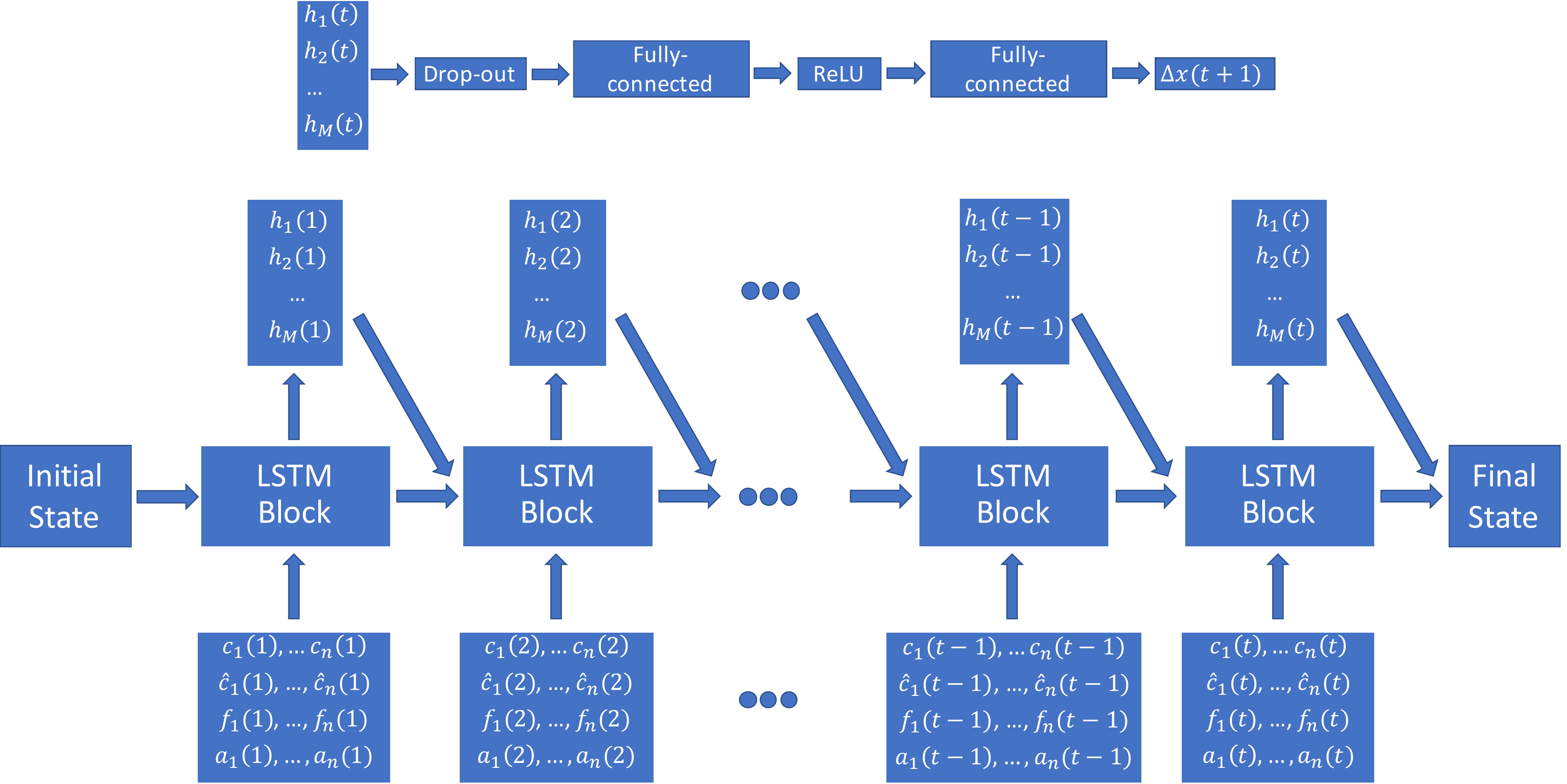}
    \caption{\small{The structure of LSTM with HHT features.}}
    \label{fig:lstm}
\end{figure}

The algorithm framework for RTE and SVM are generically applied to classification or regression tasks. The models are described in \cite{breiman2017classification,hastie2009elements} for RTE with least square boosting and in \cite{drucker1997support,vapnik2013nature} for SVM for regression. The LSTM model, on the other hand, requires design of the network structure. In this section, we describe the network structure we used for this study, which is visualized in Fig. \ref{fig:lstm}

As a type of recurrent neural network (RNN), LSTM learns the long-term dependency in sequenced data \cite{lstm_hochreiter1997long}. At each time step $t$, the LSTM block takes in new observations $c_j(t),\hat{c}_j(t), \omega_j(t),a_j(t),\ j=1,\cdots n$. The new inputs together with the network state information from the last step generate the hidden units $h_1(t),\cdots,h_M(t)$. The LSTM block and the  hidden units then propagate to the next LSTM block. For deeper networks, we apply a drop-out layer to the hidden layers  and then feed into another layer of LSTM. 

Our neural network structure consists of two layers of LSTM each with a drop-out layer. The network then links to two fully-connected layers, each with   a ReLU layer for activation. The number of hidden units in the layers are as follows:  100 - 200 for the first LSTM layer, 100 - 200 for the second LSTM layer, 50 - 100 for the first fully-connected layer, and 50 - 100 for the second fully-connected layer. For each data set, we tune the number of hidden units via cross-validation.

For the regression task to predict $y(t) = \Delta x(t+1)$,  the hidden units $h(t)$ are connected to the output $y(t)$ through a multi-layer neural network. In our study, we used multiple fully-connected layers with the activation function of rectified linear unit (ReLU). The reason to use ReLU is to create nonlinearity in the prediction and induce sparsity to choose a subset of useful signals. To train the neural network, we use the Adam solver \cite{kingma2014adam} and train multiple times with different randomization. The model with the lowest root mean square error (RMSE) on the training set is used for testing.

%Due to the nonconvexity of the loss function in the weights, global optimality is not guaranteed.

\section{Forecasting Experiments}\label{sect-exp}
In this section, we implement the machine learning forecasting experiments on the historical prices of  S\&P 500 index, CBOE volatility index (VIX), SPDR gold ETF (GLD), and the 10-year treasury yield (TNX). In the first part, we experiment feature selection for the HHT features and IMF components. The prediction follows the standard decomposition and train/test split framework. In Section \ref{sect-extra}, we discuss the extrapolation prediction and the associated end effect. The extrapolating prediction using LSTM is implemented.

\subsection{Features and IMFs Selection}
We implement the ML models using HHT features during the 10-year period 4/1/2010--3/31/2020. The data are split into a training set 4/1/2010--3/31/2018 and an unseen test set 4/1/2018--3/31/2020. The goal is short-time prediction for the next day change $\Delta x(t+1)$, based on the features up to time $t$. The comparison is between using the original time series feature $ x^{(t,\tau)}$, and with the HHT feature information $c^{(t,\tau)}_{(1,n)}, \hat{c}^{(t,\tau)}_{(1,n)}, a^{(t,\tau)}_{(1,n)}, f^{(t,\tau)}_{(1,n)}$. We set the lookback window $\tau=5$ for regression tree ensemble and SVM, while LSTM takes all the past features into the recurrent neural network, and each LSTM block controls the past information flow and memory level. We train the three ML models using four separate sets of features:
\begin{enumerate}
    \item original time series: $x(t)$,
    \item IMF components: $c_1(t),\cdots,c_n(t)$,
    \item complex IMF: $c_1(t), \hat{c}_1(t), \cdots, c_n(t),\hat{c}_n(t)$,
    \item full HHT features: $c_1(t),\hat{c}_1(t),f_1(t),a_1(t), \cdots, c_n(t),\hat{c}_n(t),f_n(t),a_n(t)$,
    \end{enumerate}

For each time series, HHT produces a large set of interpretable features corresponding to different time scales. The feature size is increased by $4n$ times, where $n$ is the number of IMF components. Through our experiments, we determine the most useful set of features for predictions using different ML methods, namely regression tree ensemble (RTE), support vector machine (SVM), and long short-term memory (LSTM) neural network. For performance comparison, we measure the mean squared errors (MSE) of the predictions over the whole horizon, based on the four financial time series S\&P 500, VIX, GLD, and 10-year treasury rate. 

Tables \ref{table:predSP500} -- \ref{table:predBond} summarize the experimental results from using   RTE, SVM, and LSTM based on  S\&P 500 (log price), VIX index, GLD (log price) and 10-year treasury rate  respectively. As benchmarks, we show the MSEs from predictions with the original time series $x$ as input feature, using the corresponding machine learning model. In all the rest parts of tables, the predictions are made using only HHT features and  do not use the original time series $x$ directly. Each row corresponds to using only a subset of HHT features, and each column corresponds to the set of IMFs  included. The first and last few components are also referred to as the high-pass (high-frequency) and low-pass (low-frequency) filters respectively.

  Tables \ref{table:predSP500}--\ref{table:predBond} show that generally the largest feature set (bottom row) lead to much reduced errors.  This suggests that HHT is capable of identifying the most important features from a time series.  Surprisingly, the feature  pair $(c, \hat{c})$, which consists of only the IMF components and the complementary imaginary parts derived from Hilbert transform in \eqref{hilbert}, yields similar or better performance than a much larger feature sets, e.g. ($c, \hat{c}, f, a$), which indicates that the most useful information in HHT is included in the complex IMFs. Considering the much smaller feature set, the complex IMFs are more preferable in terms of information criterion. Both feature sets outperform the feature set with only real IMF components, which has been explored in many previous studies of machine learning prediction with EMD.
  
    In terms of IMF selection, all three ML models perform better with all or the first few (high-frequency) IMFs than the last (low-frequency) IMFs. In many cases, the machine learning model performs best when the last IMFs are excluded (first 5). This is intuitive since low-frequency components may not be as useful as high-frequency components for predicting short-term movements, and a machine learning model with high complexity can potentially overfit on the training set with redundant information.
    
    Comparing different machine learning models, we see that LSTM does not perform as good as the other more generic models in some scenarios. This is due to the fact that the prediction framework is under the train/test split within the same decomposition. The generic models lose predictive power compared to LSTM, as we get into extrapolating prediction in the next section.

\begin{table}[ht]
\centering\begin{scriptsize}
\begin{tabular}{ l|c c c c c }
 \hline
\multicolumn{6}{c}{\textbf{Model: RTE}}   \\
\hline
Features & All IMFs & First 5 & First 4 & Last 5 & Last 4 \\
\hline
$x$ (original time series) & \multicolumn{5}{c}{2.3036} \\
\hline
$c$ & 1.2614 & 1.2185 & 1.3640 & 2.1418 & 2.2442 \\
$c, \hat{c}$ & \textbf{0.8203} & 1.0508 & 0.9540 & 2.0625 & 2.2048 \\
$c, f$ & 1.2845 & 1.1128 & 1.0747 & 2.0878 & 2.2539 \\
$c, a$ & 1.2990 & 1.1378 & 1.0581 & 2.1720 & 2.2481 \\
$c, f, a$ & 1.0137 & 1.3420 & 1.2366 & 2.0850 & 2.2892 \\
$c, \hat{c}, f, a$ & 0.8331 & 0.8713 & 1.0297 & 2.0815 & 2.2242 \\
\hline\hline
\multicolumn{6}{c}{\textbf{Model: SVM}}   \\
\hline
Features & All IMFs & First 5 & First 4 & Last 5 & Last 4 \\
\hline
$x$ (original time series) & \multicolumn{5}{c}{2.3313} \\
\hline
$c$ & 2.0456 & 2.0545 & 2.0646 & 2.2685 & 2.3087 \\
$c, \hat{c}$ & 1.9943 & 2.0214 & 1.9987 & 2.2421 & 2.2826 \\
$c, f$ & 2.1161 & \textbf{1.9825} & 2.0030 & 2.2913 & 2.3100 \\
$c, a$ & 2.1867 & 2.1490 & 2.1105 & 2.2889 & 2.3006 \\
$c, f, a$ & 2.1842 & 2.0817 & 2.1089 & 2.2953 & 2.3064 \\
$c, \hat{c}, f, a$ & 2.1446 & 2.0589 & 1.9897 & 2.2769 & 2.3023 \\
\hline\hline
\multicolumn{6}{c}{\textbf{Model: LSTM}}   \\
\hline
Features & All IMFs & First 5 & First 4 & Last 5 & Last 4 \\
\hline
$x$ (original time series) & \multicolumn{5}{c}{2.3285} \\
\hline
$c$ & 1.1400 & 1.2738 & 1.1537 & 2.0841 & 2.1510 \\
$c, \hat{c}$ & 0.6348 & 0.5938 & \textbf{0.4876} & 1.9315 & 2.1239 \\
$c, f$ & 1.2031 & 1.4041 & 1.0702 & 2.0743 & 2.1610 \\
$c, a$ & 1.4561 & 1.1516 & 1.3415 & 2.0796 & 2.2130 \\
$c, f, a$ & 1.2626 & 1.4305 & 1.1992 & 2.0707 & 2.1788 \\
$c, \hat{c}, f, a$ & 0.6834 & 0.9725 & 1.0058 & 1.9217 & 2.1233 \\
\hline
\end{tabular}\end{scriptsize}
\caption{\small{S\&P 500  prediction error (MSE, $\times 10^{-4}$) with different feature sets and components for three machine learning models (RTE (top), SVM (middle), LSTM (bottom)). The benchmark (top row for each model) is the prediction error from using the plain time series $x$ only.}}
\label{table:predSP500}
\end{table}

\begin{table}[ht]
\centering\begin{scriptsize}
\begin{tabular}{ l|c c c c c }
 \hline
\multicolumn{6}{c}{\textbf{Model: RTE}}   \\
\hline
Features & All IMFs & First 5 & First 4 & Last 5 & Last 4 \\
\hline
$x$ (original time series) & \multicolumn{5}{c}{6.2796} \\
\hline
$c$ & 2.8104 & 3.0372 & 3.9181 & 5.7601 & 6.2365 \\
$c, \hat{c}$ & 1.9488 & 1.9982 & 1.8169 & 5.6598 & 6.1334 \\
$c, f$ & 2.4096 & 2.4606 & 2.4603 & 5.7786 & 6.2631 \\
$c, a$ & 3.7021 & 3.0241 & 3.9229 & 5.8065 & 6.2387 \\
$c, f, a$ & 2.4866 & 2.7127 & 2.1297 & 5.7865 & 6.2145 \\
$c, \hat{c}, f, a$ & 2.1461 & 2.4642 & \textbf{1.5964 }& 5.5183 & 6.1631 \\
\hline\hline
\multicolumn{6}{c}{\textbf{Model: SVM}}   \\
\hline
Features & All IMFs & First 5 & First 4 & Last 5 & Last 4 \\
\hline
$x$ (original time series) & \multicolumn{5}{c}{6.4127} \\
\hline
$c$ & 5.9377 & 6.0057 & 5.6676 & 6.2427 & 6.2916 \\
$c, \hat{c}$ & 5.9248 & 5.8020 & 5.6557 & 6.1872 & 6.3019 \\
$c, f$ & 5.8231 & 5.7919 & 5.5276 & 6.2204 & 6.2988 \\
$c, a$ & 6.2414 & 6.0873 & 5.9045 & 6.2802 & 6.2950 \\
$c, f, a$ & 5.9618 & 6.0003 & 5.6598 & 6.2615 & 6.2985 \\
$c, \hat{c}, f, a$ & 5.8611 & 5.8534 & \textbf{5.3969} & 6.1940 & 6.3017 \\
\hline\hline
\multicolumn{6}{c}{\textbf{Model: LSTM}}   \\
\hline
Features & All IMFs & First 5 & First 4 & Last 5 & Last 4 \\
\hline
$x$ (original time series) & \multicolumn{5}{c}{6.3131} \\
\hline
$c$ & 3.7682 & 3.9556 & 3.4594 & 5.5977 & 6.0955 \\
$c, \hat{c}$ & 2.7004 & 2.5687 & \textbf{2.5492} & 5.3324 & 6.1218 \\
$c, f$ & 5.4949 & 5.2656 & 5.2457 & 6.4108 & 6.2548 \\
$c, a$ & 4.3976 & 4.3648 & 4.4501 & 6.2045 & 6.2070 \\
$c, f, a$ & 7.0323 & 5.7723 & 4.9104 & 6.2129 & 6.4870 \\
$c, \hat{c}, f, a$ & 4.4516 & 4.6962 & 3.9247 & 5.8894 & 6.2638 \\
\hline
\end{tabular}\end{scriptsize}
\caption{\small{VIX index prediction error (MSE) with different feature sets and components for three machine learning models (RTE (top), SVM (middle), LSTM (bottom)). The benchmark (top row for each model) is the prediction error from using the plain time series $x$ only.}}
\label{table:predVIX}
\end{table}

\begin{table}[ht]
\centering\begin{scriptsize}
\begin{tabular}{ l|c c c c c }
 \hline
\multicolumn{6}{c}{\textbf{Model: RTE}}   \\
\hline
Features & All IMFs & First 5 & First 4 & Last 5 & Last 4 \\
\hline
$x$ (original time series) & \multicolumn{5}{c}{6.9816} \\
\hline
$c$ & 2.7139 & 2.7363 & 2.5476 & 5.5666 & 6.7445 \\
$c, \hat{c}$ & 1.0644 & 1.1821 & 1.0336 & 5.2963 & 6.4202 \\
$c, f$ & 1.7681 & 1.9592 & 1.9780 & 6.0007 & 6.8471 \\
$c, a$ & 3.1483 & 2.7417 & 2.5566 & 5.8702 & 6.7719 \\
$c, f, a$ & 2.1982 & 2.0045 & 2.2453 & 5.7184 & 6.6546 \\
$c, \hat{c}, f, a$ & 0.9626 & 0.9145 & \textbf{0.8760} & 5.5237 & 6.5540 \\
\hline\hline
\multicolumn{6}{c}{\textbf{Model: SVM}}   \\
\hline
Features & All IMFs & First 5 & First 4 & Last 5 & Last 4 \\
\hline
$x$ (original time series) & \multicolumn{5}{c}{7.2381} \\
\hline
$c$ & 3.8314 & 3.8783 & 3.6813 & 6.2574 & 7.3537 \\
$c, \hat{c}$ & 2.6886 & 2.6979 & 2.7458 & 5.8944 & 6.8402 \\
$c, f$ & 3.4269 & 3.4790 & 3.3913 & 6.0189 & 6.8042 \\
$c, a$ & 4.5545 & 4.4103 & 4.3258 & 6.4397 & 7.0054 \\
$c, f, a$ & 4.3807 & 4.1316 & 3.8366 & 6.5134 & 6.8918 \\
$c, \hat{c}, f, a$ & 3.0045 & 2.6902 & \textbf{2.6795} & 6.0785 & 6.8850 \\
\hline\hline
\multicolumn{6}{c}{\textbf{Model: LSTM}}   \\
\hline
Features & All IMFs & First 5 & First 4 & Last 5 & Last 4 \\
\hline
$x$ (original time series) & \multicolumn{5}{c}{6.9792} \\
\hline
$c$ & 4.8152 & 3.4164 & 3.1468 & 5.7257 & 6.8414 \\
$c, \hat{c}$ & 1.4325 & 1.3451 & 1.4513 & 5.1071 & 6.5997 \\
$c, f$ & 2.8661 & 3.2673 & 2.8033 & 6.2854 & 6.8597 \\
$c, a$ & 3.2262 & 3.1938 & 3.5799 & 5.7604 & 6.8049 \\
$c, f, a$ & 3.2504 & 3.4774 & 3.6283 & 5.7533 & 6.7521 \\
$c, \hat{c}, f, a$ & 1.3953 & 1.5691 & \textbf{1.3248} & 5.1094 & 6.6098 \\
\hline
\end{tabular}\end{scriptsize}
\caption{\small{GLD  prediction error (MSE, $\times 10^{-5}$) with different feature sets and components for three machine learning models (RTE (top), SVM (middle), LSTM (bottom)). The benchmark (top row for each model) is the prediction error from using the plain time series $x$ only.}}
\label{table:predGLD}
\end{table}

 \begin{table}[ht]
\centering\begin{scriptsize}
\begin{tabular}{ l|c c c c c }
 \hline
\multicolumn{6}{c}{\textbf{Model: RTE}}   \\
\hline
Features & All IMFs & First 5 & First 4 & Last 5 & Last 4 \\
\hline
$x$ (original time series) & \multicolumn{5}{c}{2.6011} \\
\hline
$c$ & 1.4615 & 1.2206 & 1.1800 & 2.2609 & 2.5347 \\
$c, \hat{c}$ & \textbf{0.6244} & 0.6473 & 0.6545 & 2.1302 & 2.4472 \\
$c, f$ & 1.2036 & 1.1138 & 1.0990 & 2.2331 & 2.5236 \\
$c, a$ & 1.3131 & 1.3719 & 1.3811 & 2.2872 & 2.5832 \\
$c, f, a$ & 1.0673 & 1.0318 & 1.0657 & 2.2663 & 2.5696 \\
$c, \hat{c}, f, a$ & 0.6299 & 0.6798 & 0.8444 & 2.1907 & 2.4539 \\
\hline\hline
\multicolumn{6}{c}{\textbf{Model: SVM}}   \\
\hline
Features & All IMFs & First 5 & First 4 & Last 5 & Last 4 \\
\hline
$x$ (original time series) & \multicolumn{5}{c}{2.7631} \\
\hline
$c$ & 1.6064 & 1.5610 & 1.5968 & 2.4099 & 2.6059 \\
$c, \hat{c}$ & 1.3893 & \textbf{1.3023} & 1.3737 & 2.2918 & 2.5383 \\
$c, f$ & 1.9820 & 1.5531 & 1.4357 & 2.4251 & 2.5881 \\
$c, a$ & 1.9156 & 1.7213 & 1.6954 & 2.4124 & 2.5908 \\
$c, f, a$ & 2.0474 & 1.8175 & 1.6666 & 2.4550 & 2.5877 \\
$c, \hat{c}, f, a$ & 1.7609 & 1.4761 & 1.3948 & 2.3599 & 2.5489 \\
\hline\hline
\multicolumn{6}{c}{\textbf{Model: LSTM}}   \\
\hline
Features & All IMFs & First 5 & First 4 & Last 5 & Last 4 \\
\hline
$x$ (original time series) & \multicolumn{5}{c}{2.6071} \\
\hline
$c$ & 1.1932 & 1.1545 & 1.1304 & 2.2402 & 2.4243 \\
$c, \hat{c}$ & 0.4184 & 0.4348 & 0.5390 & 1.9995 & 2.4374 \\
$c, f$ & 1.2942 & 1.1846 & 1.1972 & 2.2446 & 2.4780 \\
$c, a$ & 1.1745 & 1.1426 & 1.0941 & 2.2503 & 2.4284 \\
$c, f, a$ & 1.2464 & 1.1732 & 1.1668 & 2.2991 & 2.5692 \\
$c, \hat{c}, f, a$ & 0.4633 & \textbf{0.3816} & 0.4190 & 2.0137 & 2.4758 \\
\hline
\end{tabular}\end{scriptsize}
\caption{\small{10-year treasury yield prediction error (MSE, $\times 10^{-3}$) with different feature sets and components for three machine learning models (RTE (top), SVM (middle), LSTM (bottom)). The benchmark (top row for each model) is the prediction error from using the plain time series $x$ only.}}
\label{table:predBond}
\end{table}

\subsection{Extrapolation and End Effect\label{sect-extra}}
As discussed in Section \ref{sect-end}, the decomposition of $x(t)$ depends on the position of $t$ in the time frame, characterized by the end effect factor $\lambda(t)$ define in \eqref{eq-lambda}. Since the distribution of the decomposed component is according to \eqref{eq-end}, depending on $\lambda$, the prediction model then becomes
\begin{equation}
x_{t+1} \sim g(c^{(t,\tau)}_{(1,n)},\hat{c}^{(t,\tau)}_{(1,n)}, a^{(t,\tau)}_{(1,n)},f^{(t,\tau)}_{(1,n)}, \lambda(t)),
\end{equation}
which means we need to incorporate $\lambda$ into the predictors. 

Furthermore, we notice that the distribution of the input and output in the time series forecasting task is not homogeneous, which requires an evolving prediction model like LSTM. The generic machine learning models RTE and SVM have poor performance on the testing error as the distribution is different from the training data. For this reason, we only show the results on LSTM with the end effect correction factor $\lambda$ as the additional feature. 

Table \ref{table:extraLSTM} shows the extrapolating prediction results with the end effect correction. We see that the different HHT feature sets all improved the prediction accuracy from using the plain time series and from the benchmark naive prediction $x_{T+1} = x(T)$. Furthermore, the complex IMF components and the full HHT feature set not only achieved lower mean square error, but also required less hidden units in the network and less training time. This shows the HHT features are better at explaining the dynamics and presenting the data.

\begin{table}[ht]
\centering\begin{small}
\begin{tabular}{ l|c c c}
 \hline
\textbf{S\&P 500} & {\textbf{Benchmark}: 5.5770}   \\
\hline
Features & MSEs & Avg. \# hidden units & Avg. train time (s) \\
\hline
$x$ (original time series) & 5.5794 & 500.0 & 6.6166 \\
$c$ & 5.0029 & 553.3 & 7.9682\\
$c, \hat{c}$ & 5.1114 & 373.1 & 5.3871\\
$c, \hat{c}, f, a$ & 5.1444 & 322.6 & 4.1803 \\
\hline
 \hline
\textbf{GLD} & {\textbf{Benchmark}: 1.5698}   \\
\hline
Features & MSEs & Avg. \# hidden units & Avg. train time (s) \\
\hline
$x$ (original time series) & 1.5781 & 550.0 & 7.7274 \\
$c$ & 1.3497 & 552.0 & 7.9372\\
$c, \hat{c}$ & 1.3304 & 348.1 & 5.4421\\
$c, \hat{c}, f, a$ & 1.3872 & 327.4 & 4.1814 \\
\hline
 \hline
\textbf{VIX} & {\textbf{Benchmark}: 87.141}   \\
\hline
Features & MSEs & Avg. \# hidden units & Avg. train time (s) \\
\hline
$x$ (original time series) & 86.828 & 400.0 & 6.8851 \\
$c$ & 79.581 & 401.0 & 5.6356\\
$c, \hat{c}$ & 79.265 & 399.5 & 5.6866\\
$c, \hat{c}, f, a$ & 75.450 & 515.7 & 8.3581 \\
\hline
 \hline
\textbf{TNX} & {\textbf{Benchmark}: 37.299}   \\
\hline
Features & MSEs & Avg. \# hidden units & Avg. train time (s) \\
\hline
$x$ (original time series) & 37.691 & 400.0 & 7.3313 \\
$c$ & 35.749 & 504.8 & 8.8135\\
$c, \hat{c}$ & 34.044 & 380.7 & 6.2190\\
$c, \hat{c}, f, a$ & 31.954 & 431.4 & 7.1112 \\
\hline
\end{tabular}\end{small}
\caption{\small{Out-of-sample prediction error (MSE, $\times 10^{-4}$) and computational cost of LSTM with different features. The benchmark is the error of naive prediction $x_{T+1} = x_T$. The first rows are the prediction with the plain time series.}}
\label{table:extraLSTM}
\end{table}

\begin{figure}[ht]
    \centering
    \includegraphics[width=0.8\textwidth]{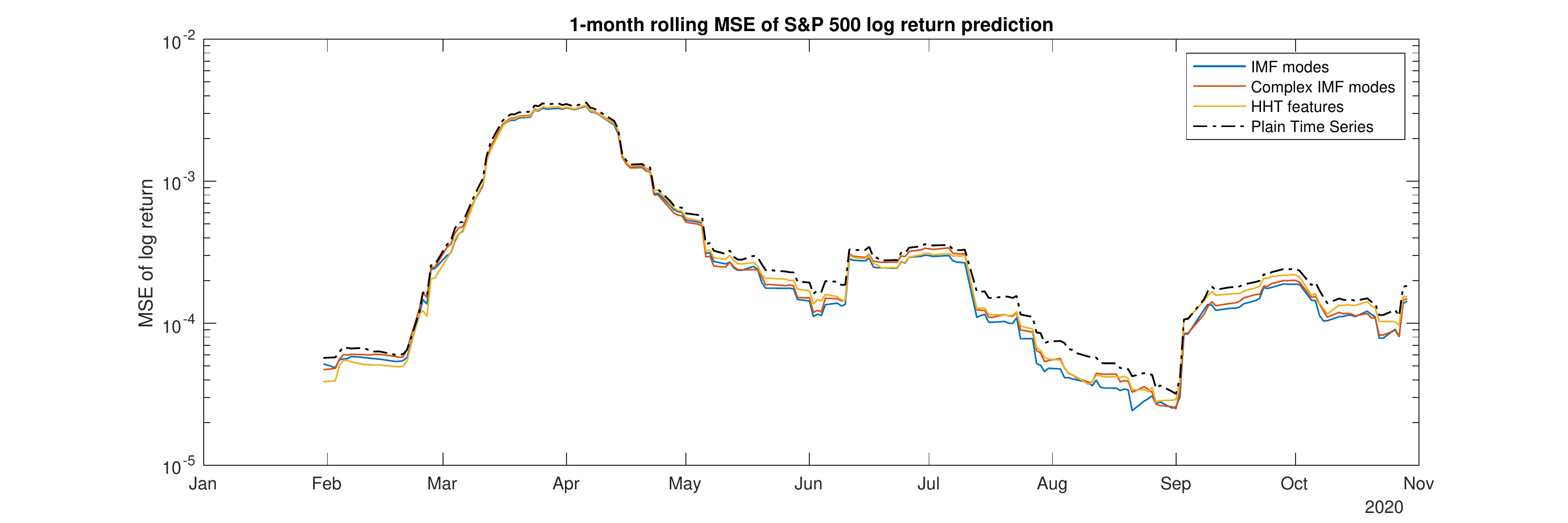}
    \includegraphics[width=0.8\textwidth]{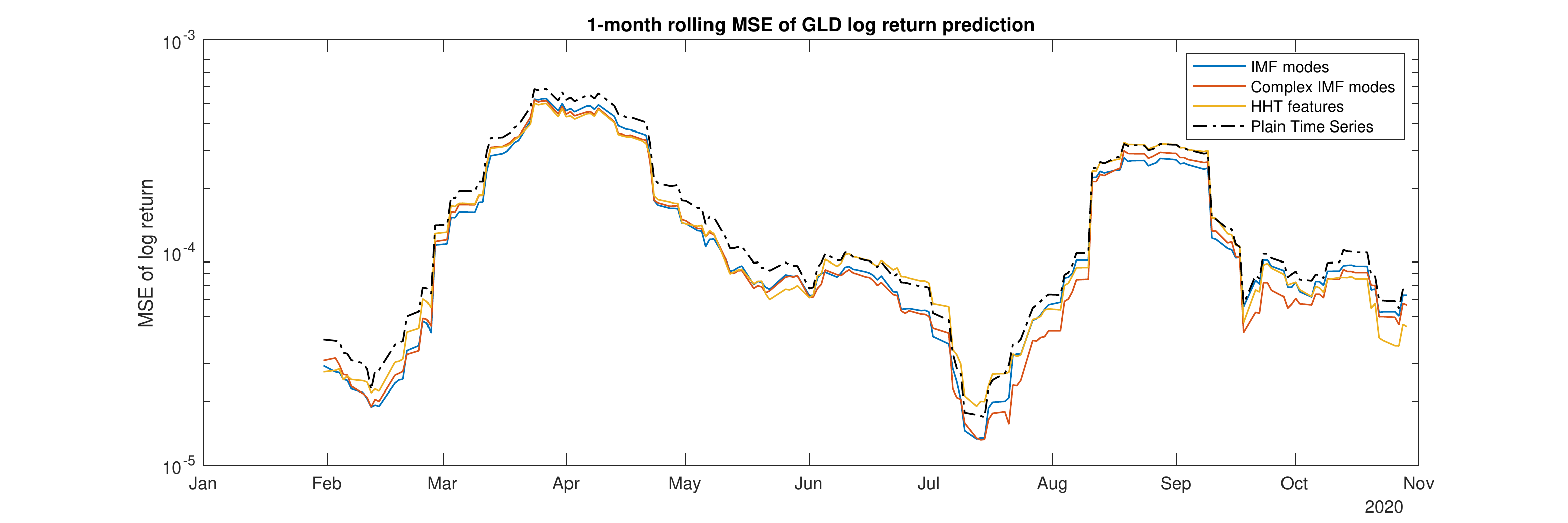}
    \includegraphics[width=0.8\textwidth]{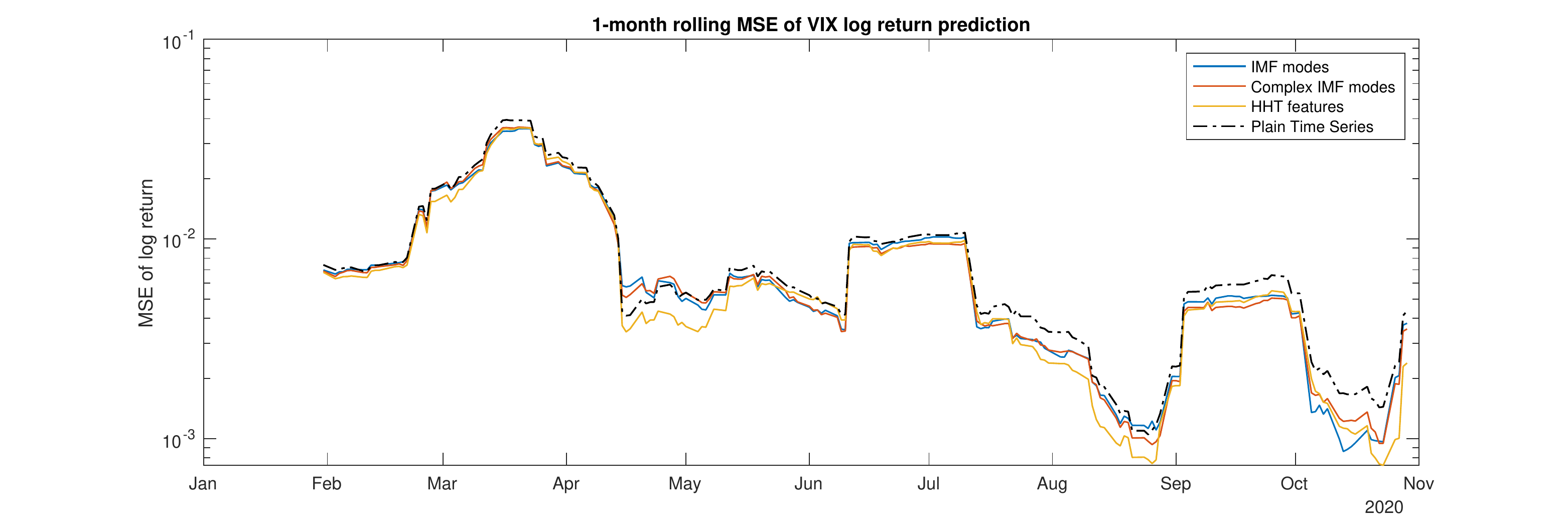}
    \includegraphics[width=0.8\textwidth]{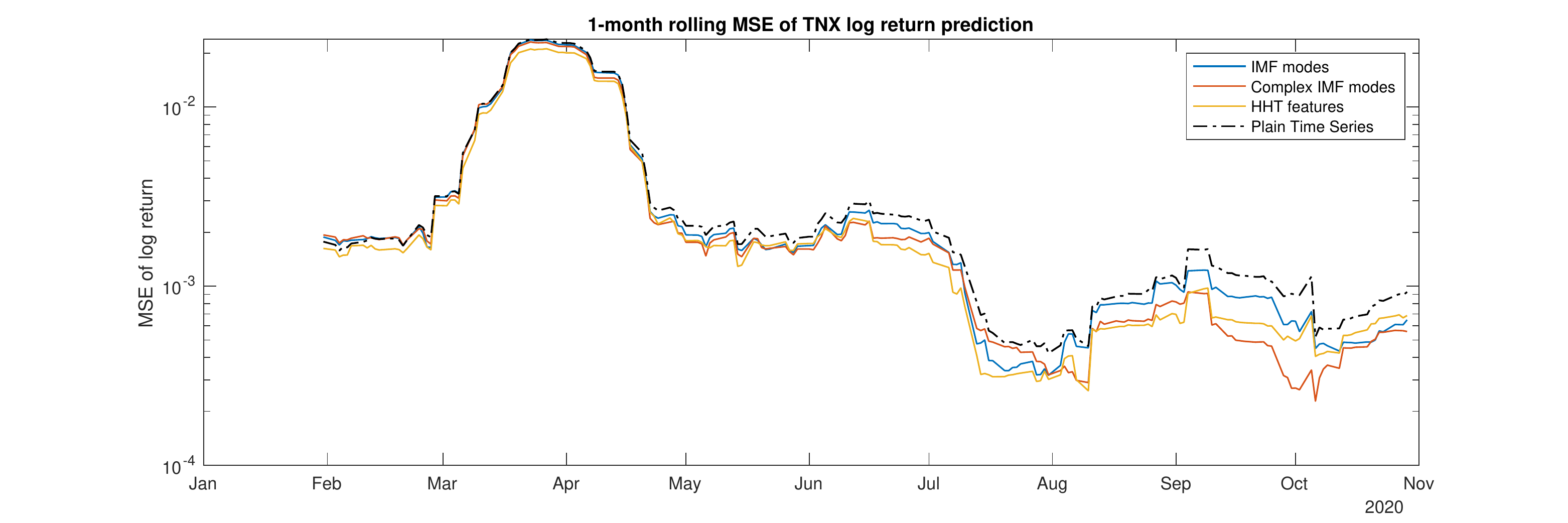}
    \caption{1-month rolling MSE of LSTM extrapolating prediction using HHT features. The benchmark (dashed line) is the prediction with plain time series as the feature.}
    \label{fig:rollingMSE}
\end{figure}

Fig. \ref{fig:rollingMSE} shows the 1-month rolling MSE of LSTM prediction using different feature sets. The experiment is implemented on S\&P 500 GLD, VIX and TNX from 1/1/2020 - 10/29/2020. The HHT feature sets consistently outperform the benchmark prediction with plain time series as feature (dashed lines). 

 \clearpage
 
 \section{Conclusion}
 We have presented the CEEMD method for multiscale analysis of  nonstationary financial time series. The key outputs of this method are the series of intrinsic mode functions, along with the   time-varying instantaneous amplitudes and instantaneous frequencies. Different combinations of modes allow us to reconstruct the time series using components of different timescales.  Using Hilbert spectral analysis, we compute  the associated instantaneous energy-frequency spectrum to illustrate the properties of various timescales embedded in the original time series. Another use of the CEEMD method is to generate a   collection of unique features that can be  integrated   into machine learning models, such as regression tree ensemble, support vector machine (SVM), and long short-term memory (LSTM) neural network. Through a series of examples with empirical financial data, we show how   HHT features can enhance the machine learning models in terms of forecasting  performance. 
\bibliographystyle{ieeetr}
\begin{small}
\bibliography{HHT_ML}
\end{small}
\end{document}